\pdfoutput=1
\documentclass[journal=jacsat,manuscript=article,maxauthors=20,etalmode=truncate,]{achemso}
\setkeys{acs}{maxauthors=20}
\setkeys{acs}{etalmode=truncate}
\usepackage[version=4]{mhchem} 
\usepackage{booktabs}
\usepackage{tabularx}
\usepackage{adjustbox}
\usepackage{rotating}
\usepackage{color,soul}
\usepackage{subcaption}
\usepackage{caption}
\usepackage{amssymb}
\usepackage{lscape}
\usepackage{graphicx}
\usepackage{amsmath}
\usepackage{comment}
\usepackage[normalem]{ulem}
\usepackage{textgreek}

\mathchardef\mhyphen="2D
\SectionNumbersOn
\author{Rohan Maniar}
\affiliation[Tulane University]
{Department of Physics and Engineering Physics, Tulane University, New Orleans, LA-70118}

\author{Priyanka B. Shukla}
\affiliation[University of Pittsburgh]
{Department of Chemical \& Petroleum Engineering, University of Pittsburgh, Pittsburgh, Pennsylvania 15261, United States}

\author{J. Karl Johnson}
\affiliation[University of Pittsburgh]
{Department of Chemical \& Petroleum Engineering, University of Pittsburgh, Pittsburgh, Pennsylvania 15261, United States}

\author{Koblar A. Jackson}
\affiliation[Central Michigan University]
{Physics Department and Science of Advanced Materials Program, Central Michigan University, Mount Pleasant,
Michigan 48859, United States}

\author{John P. Perdew}
\affiliation[Tulane University]
{Department of Physics and Engineering Physics, Tulane University, New Orleans, LA-70118}
\email{perdew@tulane.edu}

\title{Atomic Ionization: sd energy imbalance and Perdew-Zunger self-interaction correction energy penalty in 3d atoms}

\abbreviations{IR,NMR,UV}

\begin{document}





\begin{abstract}

To accurately describe the energetics of transition metal systems, density functional approximations (DFAs) must provide a balanced description of s- and d- electrons.  One measure of this is the sd transfer error, which has previously been defined as  $E(\mathrm{3d}^{n-1}\mathrm{4s}^{1}) - E(\mathrm{3d}^{n-2}\mathrm{4s}^{2})$. 
Theoretical concerns have been raised about this definition due to its evaluation of excited-state energies using ground-state DFAs. A more serious 
concern appears to be strong correlation in the 
4s$^{2}$ configuration. Here we define a ground-state
measure of the sd energy imbalance, based on the 
errors of s- and d-electron second ionization 
energies of the 3d atoms, that effectively circumvents
the aforementioned problems.  We find an improved performance as we move from LSDA to PBE to r$^{2}$SCAN for first-row transition metal atoms. However, we find large ($\sim$ 2 eV) ground-state sd energy imbalances when applying a Perdew-Zunger 1981 self-interaction correction. This is attributed to an “energy penalty” associated with the noded 3d orbitals. 
A local scaling of the self-interaction correction to LSDA results in a balance of s- and d-errors.

\end{abstract}

\section*{Introduction}
Organometallic transition-metal complexes and solids are important in many applications, including catalysis. They also present challenges to density functional theory. Necessary but not sufficient conditions on the approximation to the density functional for the exchange-correlation energy are a correct description of transition-metal atoms and atomic cations, with a correct balance between competing s and d electrons, and a correct description of the energies of transition-metal atomic cations in various charge states (regarded as proxies for oxidation states).

In the early days of Kohn-Sham density functional theory \cite{kohn1965self}, when the only standard functional for the exchange-correlation energy was the local spin density approximation (LSDA)\cite{kohn1965self}, Harris and Jones \cite{harris1978density} defined an excited-state measure of the sd transfer energy as the difference in lowest total energy between two electronic configurations of a 3d reduce atom with $n$ valence electrons,
\begin{equation}
    E(\mathrm{3d}^{n-1}\mathrm{4s}^{1}) - E(\mathrm{3d}^{n-2}\mathrm{4s}^{2})
\end{equation}
and compared their calculated values with experiments. They found that LSDA yields the correct trends across the 3d series, but seriously overbinds 3d electrons. An analysis by Gunnarsson and Jones\cite{gunnarsson1985total} found that LSDA over-stabilizes the d-rich configuration by about 1 eV relative to the s-rich configuration, and traced this error back to LSDA errors in the exchange interaction between a noded 3d orbital and the atomic core. (We will argue here for a second mechanism: strong correlation in the 
s-rich 4s$^{2}$ configuration.) They\cite{gunnarsson1985total} used accurate uniform-electron-gas exchange-correlation energies from Vosko, Wilk and Nusair \cite{vosko1980accurate}, and performed self-consistent sphericalized-orbital calculations to which they added non-spherical corrections. For each electronic configuration, they used the non-interacting or Kohn-Sham wave function of maximal possible quantum numbers $M_L$ and $M_S$,  which is a single Slater determinant. The latter choice is still used today, and in our work it is used to connect a non-interacting electronic configuration to an interacting symmetry label, but sphericalization of the orbitals is no longer used. We present the orbital configuration and not the symmetry label because the former is more natural in a discussion of sd imbalance.

Although the underlying exact Kohn-Sham theory is for the ground-state energy and density, Gunnarsson and Lundqvist \cite{gunnarsson1976exchange} proved that there is a similar exact theory for the lowest-energy state of any symmetry, although the exact exchange-correlation energy functional could be different for each symmetry. This is still a question of interest. Yang and Ayers argue that there is an exact functional of the non-interacting one-particle density matrix whose extrema are ground-state and excited-state energies.  \cite{yang2024foundation} Here we introduce a new measure of sd energy imbalance, which requires only ground-state energies. The ionization energy of any atom or ion is defined as a difference of two ground-state energies at electron numbers differing by 1. Starting from a neutral 3d atom (Sc - Zn), the first exact ionization is always the removal of an s electron, the second can be the removal of an s or a d electron with equal occurrence (suggesting a close energy competition between s and d), and the third is always the removal of a d electron. Thus, we can define a ground-state measure of sd energy imbalance as the difference between the mean errors of the 3d and 4s second ionization energies of the neutral 3d transition-metal 
atoms. We have found that the sd energy imbalance so defined decreases from 0.65 eV in LSDA \cite{kohn1965self,perdew1992accurate} to 0.44 eV in the PBE generalized gradient approximation \cite{perdew1996generalized} (GGA) to 0.09 eV in the r$^2$SCAN meta-GGA \cite{furness2020accurate}, under the increase of satisfied exact constraints on the exchange-correlation functional from 9 to 11 to 17 \cite{kaplan2023predictive}, respectively. The appropriate norms ( normally-correlated non-bonded systems to which the functional is fitted) also increase in number in this sequence. (r$^2$SCAN is a more 
computationally efficient version of SCAN, with typically
similar accuracy.)

In the neutral transition-metal atoms, the 4s shell is low enough in energy (relative to the 3d orbitals) to fill, except in Cr and Cu where the extra stability of a closed 3d subshell leads to a half-filled 4s shell.  Under ionization, the effective or Kohn-Sham potential becomes more
hydrogenic, and the 3d orbitals drop further below the 
4s, with a crossover occurring  in the +1 cations, where 
the 4s$^{1}$ and  4s$^{0}$ electronic configurations compete 
on a level playing field. For the $+n$ cations with $n \geq 2$, no 4s orbital is occupied. The standard electronic configurations are summarized in Table S1.

In our LSDA, PBE, and r$^{2}$SCAN calculations, we constrain all spin-orbital occupations to 0 or 1, because this sequence of approximations understandably approaches \cite{perdew2021spherical} the exact exchange-correlation energy defined by a constrained search \cite{levy1979universal} over wave functions, and not the one defined by a search over ensembles. Our self-consistent solutions allow spatial symmetries to break,  as is needed in some molecules and solids to capture possible strong correlation \cite{perdew2021interpretations,perdew2022symmetry,zhang2020symmetry} from approximate functionals based on normally-correlated appropriate norms. The strong correlation arises from degeneracies or near degeneracies between occupied and unoccupied spin orbitals that are removed by symmetry breaking, which transforms a strongly-correlated system into a normally-correlated one that the approximate functional can describe. The procedure defined here (if not the justification for it) is the standard way to do density functional calculations \cite{kutzler1987energies}, a procedure that leads to accurate r$^2$SCAN atomization energies for sp molecules \cite{furness2020accurate}. 

We will present further evidence from the ionization energies that the r$^2$SCAN meta-GGA, as implemented here, correctly describes isolated normally-correlated 3d neutral atoms and atomic cations without the need for a self-interaction correction. In a normally-correlated neutral atom or atomic cation, density-driven errors are expected to be small, and the functional-driven errors of r$^2$SCAN also seem to be small. 
But corrections are still needed in atomic and molecular anions where the full extra electron is not bound in r$^2$SCAN,  and in some neutral molecules and solids, where even r$^2$SCAN can make self-interaction-driven charge-transfer errors that are often reduced by a DFA+U correction as in Ref. \cite{tesch2022hubbard}. As discussed in Ref. \cite{sah2024comparison}, there are many transition-metal-compound solids in which r$^2$SCAN works well without a +U correction, and many others in which it needs such a correction, but less so \cite{sai2018evaluating} than PBE \cite{perdew1996generalized} does. Molecular dimers, including \ce{Mn_2} and \ce{Cr_2}, are also challenging due to self-interaction error, strong correlation, or both. Reference \cite{ivanov2021mn} found that the PBE binding energy curve for \ce{Mn_2} was improved by globally scaling down (by a factor of one-half) the Perdew-Zunger SIC with complex localized orbitals. Reference \cite{zhang2020localization} (without SIC) found that the realistic PBE binding energy curve of \ce{Cr_2} was degraded by r$^2$SCAN.

While our best density functional approximations (DFAs) without self-interaction correction (SIC) have relatively low ground-state measures of sd energy imbalance, size-consistent Perdew-Zunger SIC \cite{perdew1981self}, as implemented in FLOSIC-DFA\cite{pederson2014communication} using Löwdin-orthogonalized localized Fermi orbitals for the SIC, makes large sd energy imbalances. These errors are due to the d-electrons being severely under-bound in FLOSIC-DFA calculations. We hypothesize that there is an effective SIC energy penalty in the FLOSIC-DFA total energy for the noded 3d orbitals. 
While the energy penalty is present for all d-orbitals, it is maximum when the configuration is either 3d$^{5}$ or 3d$^{10}$. We show that the apparent energy penalty leads to a severe underestimation of the ionization energies across the Sc-Ge transition metal series in FLOSIC-DFA calculations when a d-electron is removed from a 3d$^{5}$ or a 3d$^{10}$ configuration. We also find that by locally scaling down the energy density of the SIC corrections, LSIC-LSDA  \cite{zope2019step} recovers from the SIC energy penalty.  (There is other evidence that full FLOSIC leads to qualitative errors in 3d systems: an excessive and unphysical symmetry-breaking in the Cr atom and dimer \cite{maniar2024symmetry}, which is eliminated by LSIC-LSDA.) 
We suspect that the reason for the SIC energy penalty is the highly lobed nature of 3d orbitals. While LSIC-LSDA can remove most of the energy penalty by scaling down the SIC corrections locally, it could be worthwhile to study these energy penalties using the complex FLOSIC method \cite{withanage2022complex} in the future.  

The Perdew-Zunger \cite{perdew1981self} SIC makes any DFA exact for all one-electron densities, but introduces errors \cite{santra2019perdew} in regions of slowly-varying density for which the DFA is typically exact. An optimal SIC is still being sought, but improved SIC can be achieved (at least for LSDA) by locally scaling down the SIC to zero in regions of slowly-varying density \cite{zope2019step}, or (at least for PBE\cite{perdew1996generalized}) by replacing real localized orbitals by complex ones \cite{klupfel2011importance}. 
A simplified version of Perdew-Zunger SIC (also used in the atomic calculations of Ref. \cite{perdew1981self}), in which the localized SIC orbitals are sphericalized canonical orbitals, was applied to ionization and transfer energies of atoms in earlier work \cite{gunnarsson1981self}. Sphericalization eliminates the angular nodes, and yields SIC-LSDA results that are close to LSDA, without the energy penalty we find here from real Fermi-L{\"o}wdin orbitals in transition-metal atoms. LSDA is less sensitive to the electron density \cite{perdew2021spherical} and density nodes \cite{sun2016communication} than are PBE and r$^2$SCAN.

We have not been able to eliminate excited states completely in this work. In a few cases, an approximate functional predicts the wrong ground-state electronic configuration (as shown later in Table \ref{tab:MLDECOMP electronic configurations}). In these cases, we have used experimental information on excitation energies to construct a reference or ``exact” ionization energy. An example is
given in Appendix A. In our LSIC-LSDA calculations, which were made on top of FLOSIC-LSDA orbitals and densities, we simply used the FLOSIC-LSDA electronic configurations of lowest FLOSIC-LSDA energy.

\section*{Methods}

\subsection*{FLOSIC formulation}
Any approximate density functional $E^{\mathrm{DFA}}[n_\uparrow,n_\downarrow]$ can be made one-electron self-interaction-free by including a Perdew-Zunger self-interaction correction (PZ-SIC) \cite{perdew1981self} defined below
\begin{equation}
E^\mathrm{PZ-SIC} [n_\uparrow,n_\downarrow] = E^\mathrm{DFA}[n_\uparrow,n_\downarrow] - \sum_{i, \sigma} (U[n_{i\sigma}] + E_\mathrm{XC} ^{\mathrm{DFA}} [n_{i\sigma},0])  \label{eqn:PZSIC-1} ,
\end{equation} 
where $U[n_{i,\sigma}]$ and $E_{\mathrm{xc}}[n_{i,\sigma},0]$ are the Hartree and exchange-correlation functionals evaluated on the single-orbital densities $n_{i,\sigma} (r)$. These densities are obtained from the energy-minimizing set of occupied one-electron orbitals \{$\phi_{i \sigma}$\} by the expression: $n_{i \sigma}(r) = |\phi_{i \sigma}(r)|^{2}$. To ensure a size-consistent theory, the single-orbital densities must be localized, a feature that is not guaranteed by the energy-minimizing orbitals themselves. It was hence suggested \cite{perdew1981self,pederson2014communication} to perform a search over restricted unitary transformations of the Kohn-Sham orbitals.

 In this work, we implement a self-interaction correction within the FLOSIC formalism \cite{pederson2014communication,pederson2015fermi,yang2017full,pederson2015self}, which, unlike previous implementations of PZ-SIC  \cite{pederson1984local,perdew1981self}, guarantees localized orbitals and a size-consistent theory. In FLOSIC, an orthonormal set of Fermi-L\"owdin orbitals (FLOs) is generated via a unitary transformation of the canonical Kohn-Sham orbitals. The unitary transformation is defined by a set of parameters called Fermi orbital descriptors (FODs), which are positions in space. The FOD positions are optimized to give optimal FLOs that minimize the PZ-SIC energy. More details on how to obtain optimal FLOs can be found in previous papers \cite{pederson2015fermi,karanovich2021electronic,shukla2023self,mishra2022study}. 
 
\subsection*{LSIC}
While FLOSIC yields functionals with improved performance for single-electron densities and stretched bonds, it deteriorates the performance for many equilibrium properties due to an overcorrection \cite{shahi2019stretched,bhattarai2021exploring,santra2019perdew}. The LSIC method, developed by Zope et al. \cite{zope2019step}, depends on locally scaling the Perdew-Zunger self-interaction correction. The LSIC correction to LSDA employs Eq. (2) with the replacements
\begin{equation}
  U^\mathrm{LSIC}[n_{i\sigma}]= \frac{1}{2} \int d^3rz_\sigma(\vec{r})n_{i\sigma}(\vec{r})   \int d^3r'\frac{n_{i\sigma}(\vec{r'})}{|\vec{r}-\vec{r'}|}   \label{eqn:LSIC C}
\end{equation}
and 
\begin{equation}
  E_\mathrm{XC}^\mathrm{LSIC}[n_{i\sigma},0] = \int d^3rz_\sigma(\vec{r})n_{i\sigma}(\vec{r}) \varepsilon_\mathrm{XC}^\mathrm{LSDA}([n_{i\sigma},0],\vec{r}), \label{eqn: LSIC XC}  
\end{equation}
In the above expressions, $z_\sigma(\vec{r}) =\frac{\tau_{\sigma}^{W} (\vec{r})}{\tau_{\sigma}(\vec{r})}$ is the scaling factor that ranges between $[0,1]$; zero for the uniform electron density and one for a one-electron density (here $\tau_{\sigma}$ is the positive kinetic energy density for occupied orbitals of spin $\sigma$, and $\tau_{\sigma}^{W}=\frac{ | \nabla n_{\sigma}|^{2}}{8 n_{\sigma}}$  ). This allows full SIE correction in one-electron-like densities and smaller corrections in slow-varying densities where the parent functional is already accurate.  $\varepsilon_\mathrm{XC}^\mathrm{DFA}$ is the density functional approximation to the local exchange-correlation energy per electron. For the results presented, we use the ``one-shot" implementation of LSIC \cite{zope2019step}. This involves evaluating the LSIC functional using the self-consistent FLOSIC-LSDA density and FOD positions.  It was shown recently that ``one-shot" LSIC results for sp molecules are very similar to those from self-consistent LSIC calculations \cite{yamamoto2023self}.

\subsection*{Computational Details}
All calculations were performed using the UTEP-NRLMOL-based FLOSIC code, which features a highly accurate numerical integration grid and extensive basis sets, inherited from its parent NRLMOL code.\cite{pederson1990variational,porezag1999optimization}. The standard NRLMOL basis set was used for all calculations described here. DFA calculations were performed using LSDA in the Perdew-Wang parameterization \cite{perdew1992accurate}, PBE \cite{perdew1996generalized}, and r$^{2}$SCAN \cite{furness2020accurate} functionals. FLOSIC calculations were performed for all functionals considered. Energy and force tolerances of $10^{-6}$ Hartree and $10^{-3}$ Hartree/Bohr,  respectively, were used. No symmetry constraints were imposed and the occupations of individual orbitals were set to integer values. For one-shot LSIC calculations, the wave functions and FOD positions of FLOSIC-LSDA calculations were used.

While the $z$-component of the expectation of the spin operator ($\mathrm{S}_{z}$) is constrained to the maximum value allowed by electronic configurations taken from NIST \cite{NIST_ASD} (Table S1), the individual numbers of s and d electrons are not separately constrained. A decomposition of the angular part of the Kohn-Sham orbitals in terms of spherical harmonics was used to analyze the calculated electronic configurations. The electronic configurations from DFA and FLOSIC-DFA calculations can be found in the Supplementary Information (Table S2 and S3). There exist cases where the computed electronic configurations do not agree with the experimental ones. In these cases, the reference experimental IP values are adjusted to be the measured ionization energy from the lowest-energy excited states consistent with the computed electronic configuration. Experimental energies for the excited states are taken from the NIST website \cite{NIST_ASD}.

FLOSIC calculations involving transition metal atoms can encounter multiple local minima for the FOD positions,\cite{kao2017self} making it important to consider multiple starting FOD configurations. In this work, we consider two sets of starting FOD positions for our FLOSIC calculations: (1) generated using FODMC \cite{schwalbe2019interpretation} and (2) reported in Kao et al. \cite{kao2017self}. We report the lower of the two converged energies.

\section*{Results and Discussion}

\subsection*{DFA and FLOSIC-DFA electronic configurations }

For the ten first-row transition metal atoms Sc-Zn, computed electronic configurations are a useful qualitative indicator of how balanced the description of s- and d-electrons is by a chosen DFA. A systematic overprediction of the total number of s- or d-electrons over experimental configurations reflects deficiencies in the underlying functionals used to compute them. In this section, we report the cases where the computed ground-state electronic configurations differ from those determined experimentally and identify key reasons for the observed mismatch.

\begin{table}[h]
	\begin{subtable}[h]{0.49\textwidth}
		\centering
  		\caption{DFA}
		\begin{tabular}{|l|l|l|}
        \hline
		   & Computed & Experiment\\
		\hline
		Fe & [Ar]3d$^{7}$4s & [Ar]3d$^{6}$4s$^{2}$\\
        \hline
		Co & [Ar]3d$^{8}$4s & [Ar]3d$^{7}$4s$^{2}$\\
        \hline
		Ni & [Ar]3d$^{9}$4s & [Ar]3d$^{8}$4s$^{2}$\\
        \hline
		Ti$^{+}$ & [Ar]3d$^{3}$ & [Ar]3d$^{2}$4s\\
        \hline
		\end{tabular}

		\label{tab: dft MLDECOMP electronic configuration}
	\end{subtable}
	\hfill
	\begin{subtable}[h]{0.49\textwidth}
		\centering
  	\caption{FLOSIC-DFA}
		\begin{tabular}{|l|l|l|}
        \hline
		   & Computed & Experiment\\
        \hline
        V$^{+}$ & [Ar]3d$^{3}$4s & [Ar] 3d$^{4}$ \\
        \hline
        Co$^{+}$ & [Ar]3d$^{7}$4s & [Ar] 3d$^{8}$ \\
        \hline
        Ni$^{+}$ & [Ar]3d$^{8}$4s & [Ar] 3d$^{9}$ \\		
        \hline
		\end{tabular}
	
		\label{tab: flosic MLDECOMP electronic configuration}
	\end{subtable}
    	\caption{Cases where the DFA (LSDA, PBE, and r$^{2}$SCAN) and FLOSIC-DFA predicted ground-state electronic configurations differ from those determined experimentally \cite{NIST_ASD}. All three DFAs predict the same ground state electronic configurations, except for Fe, for which r$^{2}$SCAN predicts the correct experimental electronic configuration. All FLOSIC-DFAs predict the same electronic configurations. }
	\label{tab:MLDECOMP electronic configurations}
\end{table}




For self-interaction-uncorrected DFAs, the asymptotic Kohn-Sham potential is known to be too repulsive, leading, for example, to positive Kohn-Sham eigenvalues in anionic systems \cite{perdew1981self}. The self-interaction error of the DFA total energy difference overbinds the 4s electron in a 4s$^{1}$ configuration, while the DFA's inability to fully describe strong correlation strongly underbinds the first 4s electron in the 4s$^{2}$ configuration. In all cases where the LSDA/PBE electronic configurations differ from experimental ones (Fe, Co, Ni, and Ti$^{+}$), the functional overestimates the number of electrons in the 3d subshell by 1, as shown in Table \ref{tab: dft MLDECOMP electronic configuration}. r$^{2}$SCAN predicts the correct electronic configuration for Fe. 

When PZ-SIC is applied, an opposite effect is observed, with a relative overstabilization of the 4s electrons compared to the 3d electrons (Table \ref{tab: flosic MLDECOMP electronic configuration}). It has previously been observed that PZ-SIC predicts higher total energies for systems with more noded occupied localized orbitals \cite{shahi2019stretched}. This introduces a relative ``SIC energy penalty" to occupying the highly-noded 3d orbitals over the less-noded 4s ones for the considered self-interaction corrected functionals.

\subsection*{Ionization energies and the sd energy imbalance} 

Ionization energies are an easy test for any DFA owing to the availability of experimental data as well as high-level wave function calculations \cite{NIST_ASD,raghavachari1989highly}. While DFA work has been done on ionization energies of first-row transition metal atoms \cite{argaman2013higher,kraisler2010ensemble}, to our knowledge none comment on the performance of the r$^{2}$SCAN meta-GGA. In this section, we report errors for the first three ionizations ($\Delta$SCF) of the 3d series (Sc-Zn) for various DFA and  FLOSIC-DFA functionals including r$^{2}$SCAN. We also use the errors in the second ionization energy to estimate the ground-state sd energy imbalances of the functionals considered. A detailed description of the individual absolute and percentage errors for all ionization processes considered in this work can be found in the Supplementary Information.
\subsubsection*{First Ionization}

For the considered atoms, experimental electronic configurations exclusively predict the removal of an electron from the 4s shell during the first ionization process. The DFA and FLOSIC-DFA electronic configurations match this expectation in most cases. Exceptions are the Fe atom in LSDA and PBE, and the Co and Ni atoms in LSDA, PBE, and r$^{2}$SCAN. 

Owing to the diffuse nature of the shell, the 4s electrons sample the long-range part of the Kohn-Sham potential. This potential has the wrong limiting form in DFA calculations and is corrected in SIC calculations \cite{perdew1981self}, suggesting that the ionization energies would be better described by self-interaction corrected functionals. Normally we find significant improvement from LSDA to PBE to r$^{2}$SCAN. However, as seen in Table \ref{tab: first ionization}, the  $\Delta$SCF results suggest only mild improvements by r$^{2}$SCAN (which has less self-interaction error than PBE) and the self-interaction-corrected functionals. We argue here that the reason for this may be the presence of strong static correlations in the fully-filled 4s$^{2}$ shell. 

Buendia et al. \cite{buendia2013quantum} have stressed the importance of accounting for strong static correlations due to the near 4s$^{2}$-4p$^{2}$ degeneracy in atoms with fully-filled 4s shells. Their calculations comparing a linear combination of two determinants (4s$^{2}$ and 4p$^{2}$ configurations) to a single determinant (4s$^{2}$ configuration) show a corresponding correlation effect on the first ionization energy, from +0.5 eV for Sc to +0.8 eV for Zn. (Starting from a two-configuration wavefunction for 4s$^2$ atoms, they obtained highly-accurate ionization energies from fixed-node Diffusion Monte Carlo.)
These correlation effects are not expected to be fully captured by standard DFAs, resulting in DFA energies that are too high for the neutral atoms with a valence 4s$^{2}$ shell, and in turn an underestimation of first ionization energies from the 4s$^{2}$ configuration. In the present work, we will present corroborating evidence that functionals predicting a 4s$^{2}$ ground state also underestimate the first ionization energy from this state because they miss its strong correlation. Similarly, symmetry-unbroken SCAN has been used \cite{perdew2022symmetry} to detect strong correlation in singlet ground-state \ce{C_2} at its equilibrium bond length.

\begin{table}[!ht]

\begin{tabular}{ lccccc}
\hline
Functional  & ME (eV) & MAE (eV) & MPE & MAPE  \\
\hline
LSDA & 0.35 & 0.53 & 4.72 & 7.24 \\ 
PBE & 0.12 & 0.45 & 1.52 & 6.38 \\
r$^{2}$SCAN & -0.14 & 0.42 & -2.07 & 5.84 \\
FLOSIC-LSDA & 0.09 & 0.19 & 1.05 & 2.39 \\
FLOSIC-PBE & -0.39 & 0.43 & -5.29 & 5.79 \\
FLOSIC-r$^{2}$SCAN & -0.31 & 0.39 & -4.33 & 5.24 \\
LSIC &  -0.31 & 0.43 & -4.06 & 5.64 \\
\hline
\end{tabular}
\caption{ Errors and percentage errors (signed and absolute) for the first ionization energy of the 3d transition metal atoms Sc-Zn. The mean errors (ME) and mean absolute errors (MAE) are reported in eV. Negative signed errors indicate that the ionization energies are underestimated. Individual errors are available in the SI.  }
\label{tab: first ionization}
\end{table}

\begin{figure}
\centering
\includegraphics[width=.8\linewidth]{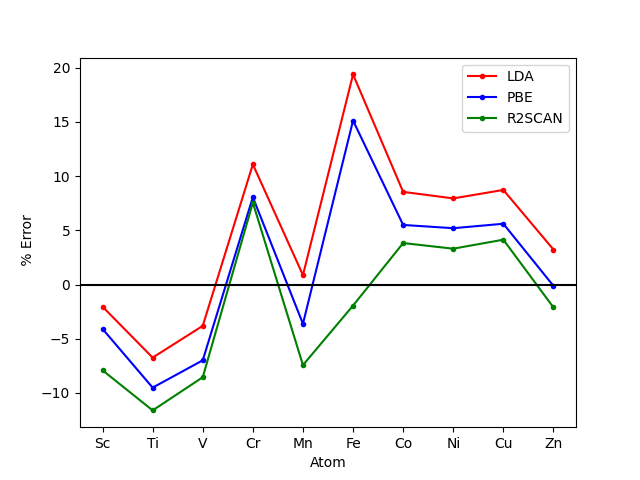}
\caption{Individual percentage errors in the first ionization energies for the 3d transition metal atoms by DFAs. The errors are most positive most positive when an electron is removed from an atom with a 4s$^1$ configuration
 (Cr, Co, Ni, Fe, and Cu for LDA and PBE, and all those 
except Fe for r$^2$SCAN ), and are small or negative  
when the electron is removed from a 4s$^2$ subshell 
(Sc, Ti, V. Mn, and Zn for LDA and PBE, and all of those 
plus Fe for r$^2$SCAN).}

\label{fig:DFT_first_IP}
\end{figure}

The individual percentage errors for the first ionization energies by self-interaction-uncorrected DFAs (Figure \ref{fig:DFT_first_IP}) further validate our hypothesis. For LSDA and PBE, the five atoms with a 3d$^{n-2}$4s$^{2}$ DFA electronic configurations (Sc, Ti, V, Mn, and Zn) have the five most negative errors. Our symmetry-unbroken r$^2$SCAN results in Fig.\ref{fig:DFT_first_IP} suggest that strong correlation in a 4s$^2$ configuration weakens from Sc and Ti to Zn. In the case of Fe, the large positive errors by LSDA/PBE can be understood as a consequence of the predicted d-removal, for which the functionals have larger errors (Table \ref{tab: sd error}). r$^{2}$SCAN predicts a 3d$^{6}$4s$^{2}$ configuration for Fe resulting in a negative error in its predicted ionization energy. Such an underestimation due to the 4s$^{2}$-4p$^{2}$ degeneracy, however, does not exist in the cases of the half-filled 4s shell. For the removal of the second 4s electron, we will see in Table \ref{tab: sd error} that r$^{2}$SCAN and the self-interaction corrected functionals perform much better than LSDA and PBE. 

\subsubsection*{Second Ionization and sd energy imbalance}
In 1978, Harris and Jones reported that LSDA calculations with sphericalized orbital densities reproduced qualitatively correct trends for the sd transfer energies \cite{harris1978density}. Following this, several investigations of the sd transfer energy were done using semilocal, hybrid, and even self-interaction-corrected functionals {\cite{johnson2017communication,gunnarsson1981self, furche2006performance,holthausen2005benchmarking, perdew1981self,harrison1983density}. However, as noted by Perdew and Furche \cite{furche2006performance}, these approaches suffer from an inability to assign total angular and spin quantum numbers ($L$ and $S$) to the Kohn-Sham determinant. (The approximate Kohn-Sham potentials of most open-shell atoms are not spherically symmetric like the external or nuclear potential, but it has been argued that, at least for sp atoms, this symmetry-breaking should have a very small effect on the calculated total energy\cite{perdew2021spherical}.)  An additional theoretical concern arises from the fact that all past explorations involve the evaluation of DFAs on excited states. We hence propose an alternate method of estimating sd energy imbalances. Of the atoms considered ($Z=21$ to 30), DFA and FLOSIC-DFA predict the removal of an s-electron for some cases and the removal of a d-electron for others, during the second ionization process (see Tables S2 and S3). We may hence define the ground-state measure of the sd energy imbalance ($\Delta_{\mathrm{sd}}$) as the difference in the errors for d- and s-removals in second ionization energies:
\begin{equation}
    \Delta_\mathrm{sd} = \mathrm{ME(d\mhyphen removal)} - \mathrm{ME(s\mhyphen removal)},
\end{equation}
where ME stands for mean errors over the 3d atoms for which a given DFA predicts the second ionization to be a given kind of electron removal. Like the error of the sd transfer energy of Eq. (1), the sd energy imbalance of Eq. (5) measures the failure of a DFA to achieve the correct balance between s and d electrons. Unlike the sd transfer errors, which are specific to individual atoms, our sd energy imbalances are statistical. We believe that this approach provides more accurate estimates of sd energy imbalance for two reasons. Firstly, it avoids the evaluation of excited-state energies. Furthermore, by only considering cases with half-filled 4s shells, it circumvents potential errors due to the large static correlations suspected for the fully-filled 4s$^{2}$ shell.  

The mean percentage errors (MPE) and mean absolute percentage errors (MAPE) for the second ionization energies are reported in Table \ref{tab: second ionization}, based upon individual d-removal and s-removal errors in Tables S11-S17. LSDA and the semi-local functionals considered consistently overestimate the ionization energies. The addition of a self-interaction correction improves the performance of LSDA but deteriorates that of the semi-local functionals, a trend that has been previously observed for other equilibrium properties \cite{shahi2019stretched,bhattarai2021exploring}. A local scaling of the self-interaction correction to LSDA provides the lowest errors of all self-interaction corrected functionals considered.

\begin{table}[!ht]

\begin{tabular}{ lccccc}
\hline
Functional  & ME (eV) & MAE (eV) & MPE & MAPE  \\
\hline
LSDA & 1.06 & 1.06 & 6.49 & 6.49 \\ 
PBE & 0.72 & 0.72 & 4.45 & 4.45 \\
r$^{2}$SCAN & 0.35 & 0.35 & 2.24 & 2.27 \\
FLOSIC-LSDA & -0.22 & 0.49 & -1.06 & 2.81 \\
FLOSIC-PBE & -0.83 & 0.83 & -4.65 & 4.65 \\
FLOSIC-r$^{2}$SCAN & -0.74 & 0.74 & -4.10 & 4.10 \\
LSIC & -0.25 & 0.28 & -1.53 & 1.75 \\
\hline
\end{tabular}
\caption{Errors and percentage errors (signed and absolute) for the second ionization energies of the 3d transition metal atoms Sc-Zn. The mean errors (ME) and mean absolute errors (MAE) are reported in eV. Negative signed errors indicate that the ionization energies are underestimated. Individual errors are available in the SI. Note the typical but not universal pattern in which errors decrease from LSDA to PBE to r$^{2}$SCAN.} 
\label{tab: second ionization}
\end{table}

Values for the ground-state measure of the sd energy imbalance are presented in Table \ref{tab: sd error}. We observe that LSDA strongly overbinds the 3d electrons compared to the 4s electrons (seen as a large positive ground-state sd energy imbalance of $\Delta_{\mathrm{sd}}=0.65$ eV), roughly in line with the findings of Harris and Jones \cite{harris1978density}. This overbinding gets reduced when one considers the semi-local functionals, with r$^{2}$SCAN displaying an sd energy imbalance under 0.1 eV. Unlike what was seen for the first ionization, here the addition of a self-interaction correction to LSDA and PBE led to significant improvement in the s-removal errors. However, the d-electrons are then severely under-bound resulting in large sd energy imbalances. A major source of these large errors is the SIC energy penalty previously mentioned. These errors are considerably reduced by locally scaling down the self-interaction correction.
\begin{table}[!ht]

\begin{tabular}{ lcccc}
\hline
Functional  & ME(d-removal) & ME(s-removal) & $\Delta_{\mathrm{sd}}$ \\
\hline
LSDA & 1.32 & 0.67 & 0.65 \\
PBE & 0.89 & 0.45 & 0.44 \\
r$^{2}$SCAN & 0.38 & 0.29 & 0.09 \\
FLOSIC-LSDA & -1.52 & 0.10 & -1.62\\
FLOSIC-PBE & -2.67 & -0.37 & -2.30 \\
FLOSIC-r$^{2}$SCAN & -2.36 & -0.33 & -2.03 \\
LSIC & -0.25 & -0.25 &  0.00 \\
\hline
\end{tabular}
\caption{ Mean errors and ground-state measures of the sd energy imbalance are reported in eV. A positive $\Delta_{\mathrm{sd}}$ implies that the DFA overbinds the 3d electrons relative to the 4s electrons. One-shot LSIC results in a cancellation of mean errors for s- and d- removals.  Individual errors are available in the SI. Note the typical but not universal pattern in which errors decrease from LSDA to PBE to r$^2$SCAN.}
\label{tab: sd error}
\end{table}

\subsubsection*{Third Ionization and nodality of the 3d orbitals}

The third ionization of first-row transition metal atoms always involves the removal of an electron from the 3d shell. From Table \ref{tab: third ionization}, we see that local and semi-local DFAs perform comparatively better on a percentage basis than in previous cases, with r$^{2}$SCAN giving the lowest percentage errors (1.27\%) of all functionals considered. While a full SIC increases the absolute percentage of errors due to the SIC energy penalty, locally scaling the self-interaction correction significantly increases the accuracy in the case of LSDA.

\begin{table}[!ht]

\begin{tabular}{ lccccc}
\hline
Functional & ME (eV) & MAE (eV) & MPE & MAPE  \\
\hline
LSDA &  1.11 & 1.11 & 3.39 & 3.39 \\
PBE & 0.77 & 0.77 & 2.42 &  2.42 \\
r$^{2}$SCAN & 0.39 & 0.39 & 1.27  & 1.27  \\
FLOSIC-LSDA & 0.55 & 1.36 & 2.17 & 4.44 \\
FLOSIC-PBE & -0.91 & 1.84 & -2.30  & 5.56 \\
FLOSIC-r$^{2}$SCAN & -0.92 & 1.25 & -2.46 & 3.69 \\
LSIC & 0.56 & 0.69 & 1.93 & 2.34 \\
\hline
\end{tabular}
\caption{ Errors and percentage errors (signed and absolute) for the third ionization of the 3d transition metal atoms Sc-Zn. The mean errors (ME) and mean absolute errors (MAE) are reported in eV. Negative signed errors indicate that the ionization energies are underestimated. Individual errors are available in the SI. Note the typical but not universal pattern in which errors decrease from LSDA to PBE to r$^2$SCAN.}
\label{tab: third ionization}
\end{table}

As the transition metal cations involved in the third ionization process only contain valence 3d electrons, we also attempt to explain the dependence of the magnitude of the SIC energy penalty on the number of electrons in the 3d subshell. The Fermi-L\"owdin orbitals are constrained to be orthogonal to one another. This directly implies that the more parallel-spin 3d electrons there are, the more noded the 3d orbitals of that spin channel should be, leading one to expect, based on arguments in Ref \cite{shahi2019stretched}, that the SIC energy penalty increases from Sc to Mn and then from Fe to Zn. 

To isolate the SIC energy contributions ($\mathrm{\Delta_{SIC}}$), Table \ref{tab: third ionization SIC penalty} presents the difference between FLOSIC-r$^{2}$SCAN and r$^{2}$SCAN@FLOSIC-r$^{2}$SCAN (the r$^{2}$SCAN functional evaluated on the FLOSIC-r$^{2}$SCAN densities) errors for the third ionization energies. While r$^{2}$SCAN@FLOSIC-r$^{2}$SCAN gives mostly small positive errors, large negative errors can be found in some cases for FLOSIC-r$^{2}$SCAN. These cases, where the SIC energy penalty is most relevant, typically occur when there are a large number of 3d electrons in the spin channel of the electron to be removed ( V to Mn and Ni to Zn ). We also note that the r$^{2}$SCAN@FLOSIC-r$^{2}$SCAN errors are very similar to those of r$^{2}$SCAN (Table S20), showing that the SIC is not changing the total density in ways that impact the calculated ionization energies. 

\begin{table}[!ht]

\begin{tabular}{ lccccc}
\hline
 & &FLOSIC- &  r$^{2}$SCAN@ & $\mathrm{\Delta_{SIC}}$ \\
  & &r$^{2}$SCAN &  FLOSIC-r$^{2}$SCAN &  \\
\hline
Sc$^{+2}$ $\rightarrow$ Sc$^{+3}$ & 3d$^{1}$ $\rightarrow$ 3d$^{0}$ & 0.63 & 0.37 & 0.25 \\
Ti$^{+2}$ $\rightarrow$ Ti$^{+3}$ & 3d$^{2}$ $\rightarrow$ 3d$^{1}$ & 0.73 & 0.57 & 0.15\\
V$^{+2}$ $\rightarrow$ V$^{+3}$ & 3d$^{3}$ $\rightarrow$ 3d$^{2}$& -1.16 & 0.52 & -1.68\\
Cr$^{+2}$ $\rightarrow$ Cr$^{+3}$ &3d$^{4}$ $\rightarrow$ 3d$^{3}$ & -1.13 & 0.34 & -1.47 \\
Mn$^{+2}$ $\rightarrow$ Mn$^{+3}$ &3d$^{5}$ $\rightarrow$ 3d$^{4}$ & -1.47 & 0.38 & -1.85\\
Fe$^{+2}$ $\rightarrow$ Fe$^{+3}$ & 3d$^{6}$ $\rightarrow$ 3d$^{5}$ & 0.08 & 0.21 & -0.13\\
Co$^{+2}$ $\rightarrow$ Co$^{+3}$ & 3d$^{7}$ $\rightarrow$ 3d$^{6}$ & 0.23 & 0.48 & -0.25\\
Ni$^{+2}$ $\rightarrow$ Ni$^{+3}$ & 3d$^{8}$ $\rightarrow$ 3d$^{7}$ & -1.90 & 0.39 & -2.29\\
Cu$^{+2}$ $\rightarrow$ Cu$^{+3}$ & 3d$^{9}$ $\rightarrow$ 3d$^{8}$ & -1.85 & -0.06 & -1.79\\
Zn$^{+2}$ $\rightarrow$ Zn$^{+3}$ & 3d$^{10}$ $\rightarrow$ 3d$^{9}$ & -3.34 & 0.11 & -3.45 \\
\hline
MAE & & 1.25& 0.34 & 1.34 \\
\hline
\end{tabular}
\caption{ Errors in eV for the indicated third ionization processes, as computed in FLOSIC-r$^{2}$SCAN, and their analysis into contributions stemming from the r$^{2}$SCAN and SIC parts of the FLOSIC total energy, respectively.  The large, negative SIC energy contributions ($\mathrm{\Delta_{SIC}}$) for Mn and Zn are seen as a consequence of the highly noded orbitals in the 3d$^{5}$ and 3d$^{10}$ configurations. The MAE for r$^2$SCAN is nearly the same in the self-consistent implementation (0.39 eV in Table 5) as on the FLOSIC-r$^2$SCAN density (0.34 eV from this table), showing that the large FLOSIC-r$^2$SCAN energy penalty is not a density-driven error.}
\label{tab: third ionization SIC penalty}
\end{table}

\subsection*{FLOSIC performance for ionization from d$^5$ or d$^{10}$ configurations}

In the previous section, we introduced the hypothesis of a SIC energy penalty associated with 3d orbitals. The results presented in Table \ref{tab: third ionization SIC penalty} suggest that the SIC part of the total energy is responsible for the penalty and that the effect is largest for filled 3d shells. We explore this hypothesis further in this section by comparing the ionization energy involving removing the first 3d electron from a filled shell (3d$^{5}$ $\rightarrow$ 3d$^{4}$  or 3d$^{10}$  $\rightarrow$ 3d$^{9}$) to the ionization energy that involves removing a second 3d electron (3d$^{4}$ $\rightarrow$ 3d$^{3}$ or 3d$^{9}$ $\rightarrow$ 3d$^{8}$). We do this by looking at the ionization energies of Mn and Zn ions and also those of highly charged Ga and Ge ions. 

Table \ref{tab: additional ionization SIC penalty} shows the error relative to experiments in the calculated ionization energies for FLOSIC-r$^2$SCAN when removing the first and second 3d electrons from a filled shell. It also decomposes the errors into contributions from the r$^2$SCAN and SIC parts of the total energy, respectively. FLOSIC-r$^2$SCAN strongly underestimates both ionization energies, but the error is clearly worse when removing the first 3d electron from a 3d$^{5}$ or 3d$^{10}$ configuration. The r$^2$SCAN part of the energy describes the ionization energies reasonably well and with roughly the same accuracy for both removals. The SIC part of the energy is therefore responsible for underestimating the ionization energies and, in all cases shown, the underestimation is worse when removing the first 3d electron. In other words, the SIC energy penalty appears greatest for a filled 3d subshell. 

\begin{table}[!ht]
\begin{tabular}{ lccccc}
\hline
 & &FLOSIC- &  r$^{2}$SCAN@ & $\mathrm{\Delta_{SIC}}$ \\
  & &r$^{2}$SCAN &  FLOSIC-r$^{2}$SCAN &  \\
\hline
Mn$^{+2}$ $\rightarrow$ Mn$^{+3}$ &3d$^{5}$ $\rightarrow$ 3d$^{4}$ &-1.47 & 0.38 & -1.85\\
Mn$^{+3}$ $\rightarrow$ Mn$^{+4}$ & 3d$^{4}$ $\rightarrow$ 3d$^{3}$ &-0.85	& 0.53   & -1.38 \\
Zn$^{+2}$ $\rightarrow$ Zn$^{+3}$ & 3d$^{10}$ $\rightarrow$ 3d$^{9}$&-3.34 & 0.11 & -3.45 \\
Zn$^{+3}$ $\rightarrow$ Zn$^{+4}$ &  3d$^{9}$ $\rightarrow$ 3d$^{8}$&-1.41 &	0.11  & -1.52 \\
Ga$^{+3}$ $\rightarrow$ Ga$^{+4}$ & 3d$^{10}$ $\rightarrow$ 3d$^{9}$&-2.58 &	0.18   & -2.76\\
Ga$^{+4}$ $\rightarrow$ Ga$^{+5}$ &  3d$^{9}$ $\rightarrow$ 3d$^{8}$&-1.82 &	0.25  & -2.07 \\
Ge$^{+4}$ $\rightarrow$ Ge$^{+5}$ & 3d$^{10}$ $\rightarrow$ 3d$^{9}$ &-2.45	& 0.25   & -2.70 \\
Ge$^{+5}$ $\rightarrow$ Ge$^{+6}$ &3d$^{9}$ $\rightarrow$ 3d$^{8}$ &-1.15 & 	0.53   & -1.68\\
\hline
MAE & & 1.88& 0.29 &2.18 \\
\hline
\end{tabular}
\caption{FLOSIC-r$^{2}$SCAN errors (in eV) relative to experimental values for the indicated ionization processes and their decomposition into contributions from the r$^{2}$SCAN (r$^{2}$SCAN@FLOSIC-r$^{2}$SCAN) and SIC ($\mathrm{\Delta_{SIC}}$) parts of the total energy. The SIC energy penalty is greatest for a d electron in a filled or half-filled shell.
}
\label{tab: additional ionization SIC penalty}
\end{table}

Tables \ref{tab:me_3d} and \ref{tab:second_3d} show error statistics for removing the first (3d$^{5}$ $\rightarrow$ 3d$^{4}$ or 3d$^{10}$ $\rightarrow$ 3d$^{9}$) and the second electron (3d$^{4}$ $\rightarrow$ 3d$^{3}$ or 3d$^{9}$ $\rightarrow$ 3d$^{8}$) from a filled 3d subshell, respectively, for all the DFAs and FLOSIC-DFAs studied here, and also for LSIC- LSDA. The uncorrected functionals overestimate both types of ionizations by similar amounts. In both cases, the error is significantly reduced from LSDA to PBE to r$^2$SCAN.

The energy for removing the first 3d electron is underestimated by all FLOSIC-DFA functionals and by FLOSIC-PBE and FLOSIC-r$^2$SCAN for removing the second 3d electron. The underestimation is much worse in the case of removing the first 3d electron. LSIC-LSDA describes the removal of the first 3d electron very accurately (MAE = 0.12 eV), but slightly overestimates the energy for the second 3d electron removal. 

Overall, the results presented in this section are consistent with the hypothesis of a FLOSIC-related energy penalty for 3d electrons that is greatest for a 3d$^{5}$ or 3d$^{10}$ configuration. 

\begin{table}[t!]
\centering
\begin{tabular}{lcccc}
\hline
Functional    & \multicolumn{1}{l}{ME (eV)} & \multicolumn{1}{l}{MAE (eV)} & \multicolumn{1}{l}{MPE} & \multicolumn{1}{l}{MAPE} \\
\hline
LSDA          & 1.32                        & 1.32                        & 4.22                    & 4.22                     \\
PBE           & 0.76                        & 0.76                        & 2.59                    & 2.59                     \\
r$^2$SCAN        & 0.32                        & 0.32                        & 1.19                    & 1.19                     \\
FLOSIC-LSDA   & -1.16                       & 1.16                        & -4.05                   & 4.05                     \\
FLOSIC-PBE    & -2.96                       & 2.96                        & -8.76                   & 8.76                     \\
FLOSIC-r$^2$SCAN & -2.43                       & 2.43                        & -7.39                   & 7.39                     \\
LSIC-LSDA     & -0.11                       & 0.12                        & -0.49                   & 0.51  \\
\hline
\end{tabular}
    \caption{Mean error (eV), mean absolute error (eV), mean percentage error (\%), and mean absolute percentage error (\%) for the removal of the first 3d electron from a 3d$^5$ or a 3d$^{10}$ configuration of Cr, Mn, Cu, Zn, Ga, and Ge atoms.}
    \label{tab:me_3d}
\end{table}

\begin{table}[t!]
\centering
\begin{tabular}{lcccc}
\hline
Functional    & \multicolumn{1}{l}{ME (eV)} & \multicolumn{1}{l}{MAE (eV)} & \multicolumn{1}{l}{MPE} & \multicolumn{1}{l}{MAPE} \\
\hline
LSDA          & 1.56                        & 1.56                        & 2.82                    & 2.82                     \\
PBE           & 0.93                        & 0.93                        & 1.69                    & 1.69                     \\
r$^2$SCAN        & 0.31                        & 0.31                        & 0.55                    & 0.55                     \\
FLOSIC-LSDA   & 0.50                         & 0.64                        & 0.59                    & 0.98                     \\
FLOSIC-PBE    & -1.58                       & 1.58                        & -3.06                   & 3.06                     \\
FLOSIC-r$^2$SCAN & -1.36                       & 1.36                        & -2.63                   & 2.63                     \\
LSIC-LSDA     & 0.78                        & 0.78                        & 1.30                     & 1.30             \\
\hline
\end{tabular}
    \caption{Mean error (eV), mean absolute error (eV), mean percentage error (\%), and mean absolute percentage error (\%) for the removal of the second 3d electron from a 3d$^5$ or a 3d$^{10}$ configuration of Cr, Mn, Cu, Zn, Ga, and Ge atoms.}
    \label{tab:second_3d}
\end{table}

\section*{Conclusion}
A correct description of transition-metal atoms and ions is a necessary but not sufficient condition for a density functional to describe the oxidation states of these ions in important complexes and solids.

In this paper, we reported DFA, FLOSIC-DFA, and one-shot LSIC electronic configurations and ionization energies for the first-row transition metal atoms. The computed electronic configurations revealed deficiencies in the various functionals considered in this work. While, from total energy differences, we see that the standard DFAs (LSDA, PBE, and r$^{2}$SCAN) overbind the 4s electron in the 4s$^{1}$ configuration, and underbind the first 4s electron in the 4s$^{2}$ configuration, the self-interaction corrected DFAs suffer from a SIC energy penalty for noded d electrons. This leads to an underestimation by DFA and an overestimation by FLOSIC-DFA of the number of 4s electrons in some cases, as shown in Table \ref{tab:MLDECOMP electronic configurations}. We briefly mention here that, at the time of writing this manuscript, we found a lower-energy solution in FLOSIC-DFA corresponding to a 3d$^{9}$4s$^{2}$ electronic configuration for the Cu atom. However, as this solution was not obtained from starting guesses provided by FODMC \cite{schwalbe2019interpretation} or Kao et al. \cite{kao2017self} (both probably biased toward NIST ground-state configurations), its energies are not reported here. 

For the first ionization process, we found that neither r$^{2}$SCAN nor self-interaction corrected functionals (except for FLOSIC-LSDA) could accurately describe the ionization energies, and postulated near-degeneracies between the fully-filled 4s$^{2}$ and 4p$^{2}$ configurations to be the reason for this. As evidence, we showed that, for the first ionization process, standard DFAs, that overestimate the 4s$^{1}$ removal energy, underestimate the first removal energy from the strongly-correlated 4s$^{2}$ ground states of the early 3d atoms. (Future work should investigate whether this strong correlation can be simulated by coaxing out a minimum-energy symmetry breaking \cite{perdew2022symmetry,perdew2021interpretations,zhang2020symmetry,maniar2024symmetry}. We did not find that here, but not finding something does not prove that it cannot be found.) The second and third ionization processes however are better described, with r$^{2}$SCAN and one-shot LSIC providing the lowest MAPE's of all functionals considered. 

The second ionization of the ten transition-metal atoms considered included both s-removals and d-removals. This allowed us to also estimate the ground-state sd energy imbalance of the functionals considered as defined in Section 3.2.2. While we found that LSDA strongly overstabilizes d-electrons compared to the s-electrons, this tendency is reduced when one considers semi-local functionals, with r$^{2}$SCAN demonstrating differences in mean signed errors for s- and d-removals that are less than 0.1 eV. The addition of a self-interaction correction in the FLOSIC scheme improves the description of s-removals but leads to a severe underbinding of the 3d electrons due to an SIC energy penalty associated with orbital lobedness, resulting in large negative ground-state sd energy imbalances. The SIC energy penalty was seen to be reduced significantly when the self-interaction correction was scaled down locally, resulting in low s-removal and d-removal errors by one-shot LSIC for the second ionization energies, and a vanishing ground-state sd energy imbalance.

For the third ionization energy, always corresponding to a d-electron removal, r$^{2}$SCAN is remarkably accurate for the ten 3d atoms Sc-Zn, with an MAE of 0.39 eV and a MAPE of 1.27 \%.

 The SIC energy penalty appears largest when an electron is removed from either a half-filled (3d$^{5}$) or completely-filled (3d$^{10}$) subshell, as evidenced by a severe underestimation of the ionization energy in these cases. This hypothesis is further supported by the similar underestimates found for the ionization energies corresponding to the removal of the first and second d-electrons from Ga$^{3+}$ and Ge$^{4+}$.
 
 The SIC energy penalty is linked to the high lobedness of the d orbitals. We observe a similar but smaller PBE-SIC energy penalty for 2p electrons 
\cite{klupfel2011importance} in closed subshells, which is strongly reduced by the use of complex localized orbitals (Fig.1 of Ref \cite{klupfel2011importance}).  
A better understanding of the penalty is important.  Predictions of physical processes that change the oxidation state of TM atoms, for example, 
could be affected. Potential improvements could involve locally scaling down the SIC, as demonstrated in this study. Another possibility is the use of complex localized orbitals (with nodeless orbital densities) to assess SIC energy terms. Complex  orbitals \cite{klupfel2011importance} 
have been shown to mitigate errors associated with the lobedness of real localized orbitals, particularly those linked to double- and triple-bonds \cite{withanage2022complex}. The same approach holds promise to eliminate the energy penalty observed in transition metal atoms discussed earlier.
We mention here that our collaborators (M.R. Pederson and K. Withanage) found that the error in the second ionization energy of the Cu atom reduced from -2.09 eV in FLOSIC-LSDA to -0.49 eV with the complex FLOSIC-LSDA (cFLOSIC) method. The result is shown in Table S34 of the Supporting Information. The cFLOSIC calculation was done by adding small imaginary components to the real FODs, resulting in complex FLOs that are optimized, keeping the density fixed, by re-optimizing the real and complex parts of the FODs. Further computational details for the cFLOSIC method can be found in Ref \cite{withanage2022complex}.

As we satisfy more exact constraints along the way from LSDA to PBE to r$^2$SCAN, we reduce but do not 
fully eliminate self-interaction error. For example, 
the large positive relative errors of first ionization
energy errors in Fig.\ref{fig:DFT_first_IP}, which we attribute to self-interaction
error in the least-bound electron of the neutral atom, reduce modestly 
from LDA to PBE to r$^2$SCAN. Fully eliminating self-interaction error can still be important, but doing so without losing the other exact constraints becomes an increasingly subtle problem. Further investigations could include complex localized orbitals for SIC, or f-electron atoms and ions, or molecules and solids, or symmetry breakings, but might better be made when and if the self-interaction correction has been further refined and the refinement has been included in the FLOSIC code.

\begin{acknowledgement}

P.B.S., J.K.J., K.A.J., and J.P.P. acknowledge
support from the U.S. Department of Energy, Office of Science, Office of Basic Energy Sciences, as part of the Computational Chemical Sciences Program, under Award No. DE--SC0018331. The work of  R.M. was supported
by the National Science Foundation under grant no, DMR-2344734 to J.P.P . Calculations were, in part, carried out at the University of Pittsburgh Center for Research Computing, RRID:SCR\_022735, using the H2P cluster, which is supported by NSF award number OAC-2117681, and the Temple EFRC computing cluster. 
We thank Professor Mark R. Pederson and Dr. Kushantha Withanage for sharing complex FLOSIC results for the Cu atom with us.  

\end{acknowledgement}

\appendix
\subsection*{Appendix A: Calculating the reference ``exact" ionization potential when the DFA ground state electronic configuration differs from the experimental electronic configuration}

There are cases where the DFA electronic configuration, computed using a decomposition of the angular part of the Kohn-Sham orbitals in terms of spherical harmonics, disagrees with the experimental electronic configuration (Table \ref{tab:MLDECOMP electronic configurations}). For such cases, we use excitation energies from \cite{NIST_ASD} to construct appropriate reference ionization energies. As an example, the reference first ionization energy for LSDA calculations of Co will now be computed.

For the Co atom, the LSDA electronic configuration ([Ar]3d$^{8}$4s from Table S2) differs from the NIST experimental configuration ([Ar]3d$^{7}$4s$^{2}$ from Table S1). From \cite{NIST_ASD}, one finds that the lowest-energy state consistent with the LSDA electronic configuration has an experimental energy 3482.82 cm$^{-1}$ (0.43 eV) above the ground state. Furthermore, the LSDA electronic configuration of Co$^{+}$ agrees with the experimental one.

Hence, the experimental ionization for Co (7.88 eV as per \cite{NIST_ASD}) needs to be reduced by 0.43 eV to give the reference ``exact" ionization potential consistent with the LSDA electronic configuration. This is the reference value used in Table S4 for the first ionization energy of Co (7.45 eV).


\bibliography{achemso-demo}

\providecommand{\latin}[1]{#1}
\makeatletter
\providecommand{\doi}
  {\begingroup\let\do\@makeother\dospecials
  \catcode`\{=1 \catcode`\}=2 \doi@aux}
\providecommand{\doi@aux}[1]{\endgroup\texttt{#1}}
\makeatother
\providecommand*\mcitethebibliography{\thebibliography}
\csname @ifundefined\endcsname{endmcitethebibliography}  {\let\endmcitethebibliography\endthebibliography}{}
\begin{mcitethebibliography}{54}
\providecommand*\natexlab[1]{#1}
\providecommand*\mciteSetBstSublistMode[1]{}
\providecommand*\mciteSetBstMaxWidthForm[2]{}
\providecommand*\mciteBstWouldAddEndPuncttrue
  {\def\EndOfBibitem{\unskip.}}
\providecommand*\mciteBstWouldAddEndPunctfalse
  {\let\EndOfBibitem\relax}
\providecommand*\mciteSetBstMidEndSepPunct[3]{}
\providecommand*\mciteSetBstSublistLabelBeginEnd[3]{}
\providecommand*\EndOfBibitem{}
\mciteSetBstSublistMode{f}
\mciteSetBstMaxWidthForm{subitem}{(\alph{mcitesubitemcount})}
\mciteSetBstSublistLabelBeginEnd
  {\mcitemaxwidthsubitemform\space}
  {\relax}
  {\relax}

\bibitem[Kohn and Sham(1965)Kohn, and Sham]{kohn1965self}
Kohn,~W.; Sham,~L.~J. Self-consistent equations including exchange and correlation effects. \emph{Physical Review} \textbf{1965}, \emph{140}, A1133\relax
\mciteBstWouldAddEndPuncttrue
\mciteSetBstMidEndSepPunct{\mcitedefaultmidpunct}
{\mcitedefaultendpunct}{\mcitedefaultseppunct}\relax
\EndOfBibitem
\bibitem[Harris and Jones(1978)Harris, and Jones]{harris1978density}
Harris,~J.; Jones,~R. Density functional theory of 3d-transition element atoms. \emph{The Journal of Chemical Physics} \textbf{1978}, \emph{68}, 3316--3317\relax
\mciteBstWouldAddEndPuncttrue
\mciteSetBstMidEndSepPunct{\mcitedefaultmidpunct}
{\mcitedefaultendpunct}{\mcitedefaultseppunct}\relax
\EndOfBibitem
\bibitem[Gunnarsson and Jones(1985)Gunnarsson, and Jones]{gunnarsson1985total}
Gunnarsson,~O.; Jones,~R. Total-energy differences: Sources of error in local-density approximations. \emph{Physical Review B} \textbf{1985}, \emph{31}, 7588\relax
\mciteBstWouldAddEndPuncttrue
\mciteSetBstMidEndSepPunct{\mcitedefaultmidpunct}
{\mcitedefaultendpunct}{\mcitedefaultseppunct}\relax
\EndOfBibitem
\bibitem[Vosko \latin{et~al.}(1980)Vosko, Wilk, and Nusair]{vosko1980accurate}
Vosko,~S.~H.; Wilk,~L.; Nusair,~M. Accurate spin-dependent electron liquid correlation energies for local spin density calculations: a critical analysis. \emph{Canadian Journal of Physics} \textbf{1980}, \emph{58}, 1200--1211\relax
\mciteBstWouldAddEndPuncttrue
\mciteSetBstMidEndSepPunct{\mcitedefaultmidpunct}
{\mcitedefaultendpunct}{\mcitedefaultseppunct}\relax
\EndOfBibitem
\bibitem[Gunnarsson and Lundqvist(1976)Gunnarsson, and Lundqvist]{gunnarsson1976exchange}
Gunnarsson,~O.; Lundqvist,~B.~I. Exchange and correlation in atoms, molecules, and solids by the spin-density-functional formalism. \emph{Physical Review B} \textbf{1976}, \emph{13}, 4274\relax
\mciteBstWouldAddEndPuncttrue
\mciteSetBstMidEndSepPunct{\mcitedefaultmidpunct}
{\mcitedefaultendpunct}{\mcitedefaultseppunct}\relax
\EndOfBibitem
\bibitem[Yang and Ayers(2024)Yang, and Ayers]{yang2024foundation}
Yang,~W.; Ayers,~P.~W. Foundation for the \textDelta SCF Approach in Density Functional Theory. \emph{arXiv preprint arXiv:2403.04604} \textbf{2024}, \relax
\mciteBstWouldAddEndPunctfalse
\mciteSetBstMidEndSepPunct{\mcitedefaultmidpunct}
{}{\mcitedefaultseppunct}\relax
\EndOfBibitem
\bibitem[Perdew and Wang(1992)Perdew, and Wang]{perdew1992accurate}
Perdew,~J.~P.; Wang,~Y. Accurate and simple analytic representation of the electron-gas correlation energy. \emph{Physical Review B} \textbf{1992}, \emph{45}, 13244\relax
\mciteBstWouldAddEndPuncttrue
\mciteSetBstMidEndSepPunct{\mcitedefaultmidpunct}
{\mcitedefaultendpunct}{\mcitedefaultseppunct}\relax
\EndOfBibitem
\bibitem[Perdew \latin{et~al.}(1996)Perdew, Burke, and Ernzerhof]{perdew1996generalized}
Perdew,~J.~P.; Burke,~K.; Ernzerhof,~M. Generalized gradient approximation made simple. \emph{Physical Review Letters} \textbf{1996}, \emph{77}, 3865\relax
\mciteBstWouldAddEndPuncttrue
\mciteSetBstMidEndSepPunct{\mcitedefaultmidpunct}
{\mcitedefaultendpunct}{\mcitedefaultseppunct}\relax
\EndOfBibitem
\bibitem[Furness \latin{et~al.}(2020)Furness, Kaplan, Ning, Perdew, and Sun]{furness2020accurate}
Furness,~J.~W.; Kaplan,~A.~D.; Ning,~J.; Perdew,~J.~P.; Sun,~J. Accurate and numerically efficient r2SCAN meta-generalized gradient approximation. \emph{The Journal of Physical Chemistry Letters} \textbf{2020}, \emph{11}, 8208--8215\relax
\mciteBstWouldAddEndPuncttrue
\mciteSetBstMidEndSepPunct{\mcitedefaultmidpunct}
{\mcitedefaultendpunct}{\mcitedefaultseppunct}\relax
\EndOfBibitem
\bibitem[Kaplan \latin{et~al.}(2023)Kaplan, Levy, and Perdew]{kaplan2023predictive}
Kaplan,~A.~D.; Levy,~M.; Perdew,~J.~P. The predictive power of exact constraints and appropriate norms in density functional theory. \emph{Annual Review of Physical Chemistry} \textbf{2023}, \emph{74}, 193--218\relax
\mciteBstWouldAddEndPuncttrue
\mciteSetBstMidEndSepPunct{\mcitedefaultmidpunct}
{\mcitedefaultendpunct}{\mcitedefaultseppunct}\relax
\EndOfBibitem
\bibitem[Chowdhury and Perdew(2021)Chowdhury, and Perdew]{perdew2021spherical}
Chowdhury,~S. T. u.~R.; Perdew,~J.~P. {Spherical vs non-spherical and symmetry-preserving vs symmetry-breaking densities of open-shell atoms in density functional theory}. \emph{The Journal of Chemical Physics} \textbf{2021}, \emph{155}, 234110\relax
\mciteBstWouldAddEndPuncttrue
\mciteSetBstMidEndSepPunct{\mcitedefaultmidpunct}
{\mcitedefaultendpunct}{\mcitedefaultseppunct}\relax
\EndOfBibitem
\bibitem[Levy(1979)]{levy1979universal}
Levy,~M. Universal variational functionals of electron densities, first-order density matrices, and natural spin-orbitals and solution of the v-representability problem. \emph{Proceedings of the National Academy of Sciences} \textbf{1979}, \emph{76}, 6062--6065\relax
\mciteBstWouldAddEndPuncttrue
\mciteSetBstMidEndSepPunct{\mcitedefaultmidpunct}
{\mcitedefaultendpunct}{\mcitedefaultseppunct}\relax
\EndOfBibitem
\bibitem[Perdew \latin{et~al.}(2021)Perdew, Ruzsinszky, Sun, Nepal, and Kaplan]{perdew2021interpretations}
Perdew,~J.~P.; Ruzsinszky,~A.; Sun,~J.; Nepal,~N.~K.; Kaplan,~A.~D. Interpretations of ground-state symmetry breaking and strong correlation in wavefunction and density functional theories. \emph{Proceedings of the National Academy of Sciences} \textbf{2021}, \emph{118}, e2017850118\relax
\mciteBstWouldAddEndPuncttrue
\mciteSetBstMidEndSepPunct{\mcitedefaultmidpunct}
{\mcitedefaultendpunct}{\mcitedefaultseppunct}\relax
\EndOfBibitem
\bibitem[Perdew \latin{et~al.}(2022)Perdew, Chowdhury, Shahi, Kaplan, Song, and Bylaska]{perdew2022symmetry}
Perdew,~J.~P.; Chowdhury,~S. T. U.~R.; Shahi,~C.; Kaplan,~A.~D.; Song,~D.; Bylaska,~E.~J. Symmetry breaking with the SCAN density functional describes strong correlation in the singlet carbon dimer. \emph{The Journal of Physical Chemistry A} \textbf{2022}, \emph{127}, 384--389\relax
\mciteBstWouldAddEndPuncttrue
\mciteSetBstMidEndSepPunct{\mcitedefaultmidpunct}
{\mcitedefaultendpunct}{\mcitedefaultseppunct}\relax
\EndOfBibitem
\bibitem[Zhang \latin{et~al.}(2020)Zhang, Furness, Zhang, Wang, Zunger, and Sun]{zhang2020symmetry}
Zhang,~Y.; Furness,~J.; Zhang,~R.; Wang,~Z.; Zunger,~A.; Sun,~J. Symmetry-breaking polymorphous descriptions for correlated materials without interelectronic U. \emph{Physical Review B} \textbf{2020}, \emph{102}, 045112\relax
\mciteBstWouldAddEndPuncttrue
\mciteSetBstMidEndSepPunct{\mcitedefaultmidpunct}
{\mcitedefaultendpunct}{\mcitedefaultseppunct}\relax
\EndOfBibitem
\bibitem[Kutzler and Painter(1987)Kutzler, and Painter]{kutzler1987energies}
Kutzler,~F.~W.; Painter,~G. Energies of atoms with nonspherical charge densities calculated with nonlocal density-functional theory. \emph{Physical Review Letters} \textbf{1987}, \emph{59}, 1285\relax
\mciteBstWouldAddEndPuncttrue
\mciteSetBstMidEndSepPunct{\mcitedefaultmidpunct}
{\mcitedefaultendpunct}{\mcitedefaultseppunct}\relax
\EndOfBibitem
\bibitem[Tesch and Kowalski(2022)Tesch, and Kowalski]{tesch2022hubbard}
Tesch,~R.; Kowalski,~P.~M. Hubbard U parameters for transition metals from first principles. \emph{Physical Review B} \textbf{2022}, \emph{105}, 195153\relax
\mciteBstWouldAddEndPuncttrue
\mciteSetBstMidEndSepPunct{\mcitedefaultmidpunct}
{\mcitedefaultendpunct}{\mcitedefaultseppunct}\relax
\EndOfBibitem
\bibitem[Sah \latin{et~al.}(2024)Sah, Zdilla, Borguet, and Perdew]{sah2024comparison}
Sah,~R.~K.; Zdilla,~M.~J.; Borguet,~E.; Perdew,~J.~P. Comparing meta-GGAs, DFT+ U corrections, and hybrid functionals for polaronic point defects in layered \ce{MnO_2}, \ce{NiO_2} and \ce{KCoO_2}. \emph{Physical Review B} \textbf{2024}, \emph{110}, 184114\relax
\mciteBstWouldAddEndPuncttrue
\mciteSetBstMidEndSepPunct{\mcitedefaultmidpunct}
{\mcitedefaultendpunct}{\mcitedefaultseppunct}\relax
\EndOfBibitem
\bibitem[Sai~Gautam and Carter(2018)Sai~Gautam, and Carter]{sai2018evaluating}
Sai~Gautam,~G.; Carter,~E.~A. Evaluating transition metal oxides within DFT-SCAN and SCAN+ U frameworks for solar thermochemical applications. \emph{Physical Review Materials} \textbf{2018}, \emph{2}, 095401\relax
\mciteBstWouldAddEndPuncttrue
\mciteSetBstMidEndSepPunct{\mcitedefaultmidpunct}
{\mcitedefaultendpunct}{\mcitedefaultseppunct}\relax
\EndOfBibitem
\bibitem[Ivanov \latin{et~al.}(2021)Ivanov, Ghosh, Jonsson, and Jonsson]{ivanov2021mn}
Ivanov,~A.~V.; Ghosh,~T.~K.; Jonsson,~E.~O.; Jonsson,~H. Mn dimer can be described accurately with density functional calculations when self-interaction correction is applied. \emph{The journal of physical chemistry letters} \textbf{2021}, \emph{12}, 4240--4246\relax
\mciteBstWouldAddEndPuncttrue
\mciteSetBstMidEndSepPunct{\mcitedefaultmidpunct}
{\mcitedefaultendpunct}{\mcitedefaultseppunct}\relax
\EndOfBibitem
\bibitem[Zhang \latin{et~al.}(2020)Zhang, Zhang, and Singh]{zhang2020localization}
Zhang,~Y.; Zhang,~W.; Singh,~D.~J. Localization in the SCAN meta-generalized gradient approximation functional leading to broken symmetry ground states for graphene and benzene. \emph{Physical Chemistry Chemical Physics} \textbf{2020}, \emph{22}, 19585--19591\relax
\mciteBstWouldAddEndPuncttrue
\mciteSetBstMidEndSepPunct{\mcitedefaultmidpunct}
{\mcitedefaultendpunct}{\mcitedefaultseppunct}\relax
\EndOfBibitem
\bibitem[Perdew and Zunger(1981)Perdew, and Zunger]{perdew1981self}
Perdew,~J.~P.; Zunger,~A. Self-interaction correction to density-functional approximations for many-electron systems. \emph{Physical Review B} \textbf{1981}, \emph{23}, 5048\relax
\mciteBstWouldAddEndPuncttrue
\mciteSetBstMidEndSepPunct{\mcitedefaultmidpunct}
{\mcitedefaultendpunct}{\mcitedefaultseppunct}\relax
\EndOfBibitem
\bibitem[Pederson \latin{et~al.}(2014)Pederson, Ruzsinszky, and Perdew]{pederson2014communication}
Pederson,~M.~R.; Ruzsinszky,~A.; Perdew,~J.~P. Communication: Self-interaction correction with unitary invariance in density functional theory. \emph{The Journal of Chemical Physics} \textbf{2014}, \emph{140}\relax
\mciteBstWouldAddEndPuncttrue
\mciteSetBstMidEndSepPunct{\mcitedefaultmidpunct}
{\mcitedefaultendpunct}{\mcitedefaultseppunct}\relax
\EndOfBibitem
\bibitem[Zope \latin{et~al.}(2019)Zope, Yamamoto, Diaz, Baruah, Peralta, Jackson, Santra, and Perdew]{zope2019step}
Zope,~R.~R.; Yamamoto,~Y.; Diaz,~C.~M.; Baruah,~T.; Peralta,~J.~E.; Jackson,~K.~A.; Santra,~B.; Perdew,~J.~P. A step in the direction of resolving the paradox of Perdew-Zunger self-interaction correction. \emph{The Journal of Chemical Physics} \textbf{2019}, \emph{151}, 214108\relax
\mciteBstWouldAddEndPuncttrue
\mciteSetBstMidEndSepPunct{\mcitedefaultmidpunct}
{\mcitedefaultendpunct}{\mcitedefaultseppunct}\relax
\EndOfBibitem
\bibitem[Maniar \latin{et~al.}(2024)Maniar, Withanage, Shahi, Kaplan, Perdew, and Pederson]{maniar2024symmetry}
Maniar,~R.; Withanage,~K.~P.; Shahi,~C.; Kaplan,~A.~D.; Perdew,~J.~P.; Pederson,~M.~R. Symmetry breaking and self-interaction correction in the chromium atom and dimer. \emph{The Journal of Chemical Physics} \textbf{2024}, \emph{160}\relax
\mciteBstWouldAddEndPuncttrue
\mciteSetBstMidEndSepPunct{\mcitedefaultmidpunct}
{\mcitedefaultendpunct}{\mcitedefaultseppunct}\relax
\EndOfBibitem
\bibitem[Withanage \latin{et~al.}(2022)Withanage, Jackson, and Pederson]{withanage2022complex}
Withanage,~K.~P.; Jackson,~K.~A.; Pederson,~M.~R. Complex Fermi--L{\"o}wdin orbital self-interaction correction. \emph{The Journal of Chemical Physics} \textbf{2022}, \emph{156}, 231103\relax
\mciteBstWouldAddEndPuncttrue
\mciteSetBstMidEndSepPunct{\mcitedefaultmidpunct}
{\mcitedefaultendpunct}{\mcitedefaultseppunct}\relax
\EndOfBibitem
\bibitem[Santra and Perdew(2019)Santra, and Perdew]{santra2019perdew}
Santra,~B.; Perdew,~J.~P. Perdew-Zunger self-interaction correction: How wrong for uniform densities and large-\ce{Z} atoms? \emph{The Journal of Chemical Physics} \textbf{2019}, \emph{150}\relax
\mciteBstWouldAddEndPuncttrue
\mciteSetBstMidEndSepPunct{\mcitedefaultmidpunct}
{\mcitedefaultendpunct}{\mcitedefaultseppunct}\relax
\EndOfBibitem
\bibitem[Kl{\"u}pfel \latin{et~al.}(2011)Kl{\"u}pfel, Kl{\"u}pfel, and J{\'o}nsson]{klupfel2011importance}
Kl{\"u}pfel,~S.; Kl{\"u}pfel,~P.; J{\'o}nsson,~H. Importance of complex orbitals in calculating the self-interaction-corrected ground state of atoms. \emph{Physical Review A} \textbf{2011}, \emph{84}, 050501\relax
\mciteBstWouldAddEndPuncttrue
\mciteSetBstMidEndSepPunct{\mcitedefaultmidpunct}
{\mcitedefaultendpunct}{\mcitedefaultseppunct}\relax
\EndOfBibitem
\bibitem[Gunnarsson and Jones(1981)Gunnarsson, and Jones]{gunnarsson1981self}
Gunnarsson,~O.; Jones,~R. Self-interaction corrections in the density functional formalism. \emph{Solid State Communications} \textbf{1981}, \emph{37}, 249--252\relax
\mciteBstWouldAddEndPuncttrue
\mciteSetBstMidEndSepPunct{\mcitedefaultmidpunct}
{\mcitedefaultendpunct}{\mcitedefaultseppunct}\relax
\EndOfBibitem
\bibitem[Sun \latin{et~al.}(2016)Sun, Perdew, Yang, and Peng]{sun2016communication}
Sun,~J.; Perdew,~J.~P.; Yang,~Z.; Peng,~H. Communication: Near-locality of exchange and correlation density functionals for 1-and 2-electron systems. \emph{The Journal of Chemical Physics} \textbf{2016}, \emph{144}\relax
\mciteBstWouldAddEndPuncttrue
\mciteSetBstMidEndSepPunct{\mcitedefaultmidpunct}
{\mcitedefaultendpunct}{\mcitedefaultseppunct}\relax
\EndOfBibitem
\bibitem[Pederson(2015)]{pederson2015fermi}
Pederson,~M.~R. Fermi Orbital Derivatives in Self-Interaction Corrected Density Functional Theory: Applications to Closed Shell Atoms. \emph{J. Chem. Phys.} \textbf{2015}, \emph{142}, 064112\relax
\mciteBstWouldAddEndPuncttrue
\mciteSetBstMidEndSepPunct{\mcitedefaultmidpunct}
{\mcitedefaultendpunct}{\mcitedefaultseppunct}\relax
\EndOfBibitem
\bibitem[Yang \latin{et~al.}(2017)Yang, Pederson, and Perdew]{yang2017full}
Yang,~Z.-h.; Pederson,~M.~R.; Perdew,~J.~P. Full self-consistency in the Fermi-orbital self-interaction correction. \emph{Physical Review A} \textbf{2017}, \emph{95}, 052505\relax
\mciteBstWouldAddEndPuncttrue
\mciteSetBstMidEndSepPunct{\mcitedefaultmidpunct}
{\mcitedefaultendpunct}{\mcitedefaultseppunct}\relax
\EndOfBibitem
\bibitem[Pederson and Baruah(2015)Pederson, and Baruah]{pederson2015self}
Pederson,~M.~R.; Baruah,~T. \emph{Advances in Atomic, Molecular, and Optical Physics}; Elsevier: Amsterdam, 2015; Vol.~64; pp 153--180\relax
\mciteBstWouldAddEndPuncttrue
\mciteSetBstMidEndSepPunct{\mcitedefaultmidpunct}
{\mcitedefaultendpunct}{\mcitedefaultseppunct}\relax
\EndOfBibitem
\bibitem[Pederson \latin{et~al.}(1984)Pederson, Heaton, and Lin]{pederson1984local}
Pederson,~M.~R.; Heaton,~R.~A.; Lin,~C.~C. Local-density Hartree--Fock theory of electronic states of molecules with self-interaction correction. \emph{The Journal of Chemical Physics} \textbf{1984}, \emph{80}, 1972--1975\relax
\mciteBstWouldAddEndPuncttrue
\mciteSetBstMidEndSepPunct{\mcitedefaultmidpunct}
{\mcitedefaultendpunct}{\mcitedefaultseppunct}\relax
\EndOfBibitem
\bibitem[Karanovich \latin{et~al.}(2021)Karanovich, Yamamoto, Jackson, and Park]{karanovich2021electronic}
Karanovich,~A.; Yamamoto,~Y.; Jackson,~K.~A.; Park,~K. Electronic structure of mononuclear \ce{Cu}-based molecule from density-functional theory with self-interaction correction. \emph{The Journal of Chemical Physics} \textbf{2021}, \emph{155}, 014106\relax
\mciteBstWouldAddEndPuncttrue
\mciteSetBstMidEndSepPunct{\mcitedefaultmidpunct}
{\mcitedefaultendpunct}{\mcitedefaultseppunct}\relax
\EndOfBibitem
\bibitem[Shukla \latin{et~al.}(2023)Shukla, Mishra, Baruah, Zope, Jackson, and Johnson]{shukla2023self}
Shukla,~P.~B.; Mishra,~P.; Baruah,~T.; Zope,~R.~R.; Jackson,~K.~A.; Johnson,~J.~K. How Do Self-Interaction Errors Associated with Stretched Bonds Affect Barrier Height Predictions? \emph{The Journal of Physical Chemistry A} \textbf{2023}, \emph{127}, 1750--1759\relax
\mciteBstWouldAddEndPuncttrue
\mciteSetBstMidEndSepPunct{\mcitedefaultmidpunct}
{\mcitedefaultendpunct}{\mcitedefaultseppunct}\relax
\EndOfBibitem
\bibitem[Mishra \latin{et~al.}(2022)Mishra, Yamamoto, Johnson, Jackson, Zope, and Baruah]{mishra2022study}
Mishra,~P.; Yamamoto,~Y.; Johnson,~J.~K.; Jackson,~K.~A.; Zope,~R.~R.; Baruah,~T. Study of self-interaction-errors in barrier heights using locally scaled and Perdew--Zunger self-interaction methods. \emph{The Journal of Chemical Physics} \textbf{2022}, \emph{156}, 014306\relax
\mciteBstWouldAddEndPuncttrue
\mciteSetBstMidEndSepPunct{\mcitedefaultmidpunct}
{\mcitedefaultendpunct}{\mcitedefaultseppunct}\relax
\EndOfBibitem
\bibitem[Shahi \latin{et~al.}(2019)Shahi, Bhattarai, Wagle, Santra, Schwalbe, Hahn, Kortus, Jackson, Peralta, Trepte, Lehtola, Nepal, Myneni, Neupane, Adhikari, Ruzsinszky, Yamamoto, Baruah, Zope, and Perdew]{shahi2019stretched}
Shahi,~C.; Bhattarai,~P.; Wagle,~K.; Santra,~B.; Schwalbe,~S.; Hahn,~T.; Kortus,~J.; Jackson,~K.~A.; Peralta,~J.~E.; Trepte,~K.; Lehtola,~S.; Nepal,~N.~K.; Myneni,~H.; Neupane,~B.; Adhikari,~S.; Ruzsinszky,~A.; Yamamoto,~Y.; Baruah,~T.; Zope,~R.~R.; Perdew,~J.~P. Stretched or noded orbital densities and self-interaction correction in density functional theory. \emph{The Journal of Chemical Physics} \textbf{2019}, \emph{150}, 174102\relax
\mciteBstWouldAddEndPuncttrue
\mciteSetBstMidEndSepPunct{\mcitedefaultmidpunct}
{\mcitedefaultendpunct}{\mcitedefaultseppunct}\relax
\EndOfBibitem
\bibitem[Bhattarai \latin{et~al.}(2021)Bhattarai, Santra, Wagle, Yamamoto, Zope, Ruzsinszky, Jackson, and Perdew]{bhattarai2021exploring}
Bhattarai,~P.; Santra,~B.; Wagle,~K.; Yamamoto,~Y.; Zope,~R.~R.; Ruzsinszky,~A.; Jackson,~K.~A.; Perdew,~J.~P. Exploring and enhancing the accuracy of interior-scaled Perdew--Zunger self-interaction correction. \emph{The Journal of Chemical Physics} \textbf{2021}, \emph{154}, 094105\relax
\mciteBstWouldAddEndPuncttrue
\mciteSetBstMidEndSepPunct{\mcitedefaultmidpunct}
{\mcitedefaultendpunct}{\mcitedefaultseppunct}\relax
\EndOfBibitem
\bibitem[Yamamoto \latin{et~al.}(2023)Yamamoto, Baruah, Chang, Romero, and Zope]{yamamoto2023self}
Yamamoto,~Y.; Baruah,~T.; Chang,~P.-H.; Romero,~S.; Zope,~R.~R. Self-consistent implementation of locally scaled self-interaction-correction method. \emph{The Journal of Chemical Physics} \textbf{2023}, \emph{158}, 064114\relax
\mciteBstWouldAddEndPuncttrue
\mciteSetBstMidEndSepPunct{\mcitedefaultmidpunct}
{\mcitedefaultendpunct}{\mcitedefaultseppunct}\relax
\EndOfBibitem
\bibitem[Pederson and Jackson(1990)Pederson, and Jackson]{pederson1990variational}
Pederson,~M.~R.; Jackson,~K.~A. Variational mesh for quantum-mechanical simulations. \emph{Physical Review B} \textbf{1990}, \emph{41}, 7453\relax
\mciteBstWouldAddEndPuncttrue
\mciteSetBstMidEndSepPunct{\mcitedefaultmidpunct}
{\mcitedefaultendpunct}{\mcitedefaultseppunct}\relax
\EndOfBibitem
\bibitem[Porezag and Pederson(1999)Porezag, and Pederson]{porezag1999optimization}
Porezag,~D.; Pederson,~M.~R. Optimization of Gaussian basis sets for density-functional calculations. \emph{Physical Review A} \textbf{1999}, \emph{60}, 2840\relax
\mciteBstWouldAddEndPuncttrue
\mciteSetBstMidEndSepPunct{\mcitedefaultmidpunct}
{\mcitedefaultendpunct}{\mcitedefaultseppunct}\relax
\EndOfBibitem
\bibitem[Kramida \latin{et~al.}(2023)Kramida, {Yu.~Ralchenko}, Reader, and {and NIST ASD Team}]{NIST_ASD}
Kramida,~A.; {Yu.~Ralchenko}; Reader,~J.; {and NIST ASD Team} {NIST Atomic Spectra Database (ver. 5.11), [Online]. Available: {\tt{https://physics.nist.gov/asd}} [2016, January 31]. National Institute of Standards and Technology, Gaithersburg, MD.}, 2023\relax
\mciteBstWouldAddEndPuncttrue
\mciteSetBstMidEndSepPunct{\mcitedefaultmidpunct}
{\mcitedefaultendpunct}{\mcitedefaultseppunct}\relax
\EndOfBibitem
\bibitem[Kao \latin{et~al.}(2017)Kao, Withanage, Hahn, Batool, Kortus, and Jackson]{kao2017self}
Kao,~D.-y.; Withanage,~K.; Hahn,~T.; Batool,~J.; Kortus,~J.; Jackson,~K. Self-consistent self-interaction corrected density functional theory calculations for atoms using Fermi-L{\"o}wdin orbitals: Optimized Fermi-orbital descriptors for \ce{Li}--\ce{Kr}. \emph{The Journal of Chemical Physics} \textbf{2017}, \emph{147}, 164107\relax
\mciteBstWouldAddEndPuncttrue
\mciteSetBstMidEndSepPunct{\mcitedefaultmidpunct}
{\mcitedefaultendpunct}{\mcitedefaultseppunct}\relax
\EndOfBibitem
\bibitem[Schwalbe \latin{et~al.}(2019)Schwalbe, Trepte, Fiedler, Johnson, Kraus, Hahn, Peralta, Jackson, and Kortus]{schwalbe2019interpretation}
Schwalbe,~S.; Trepte,~K.; Fiedler,~L.; Johnson,~A.~I.; Kraus,~J.; Hahn,~T.; Peralta,~J.~E.; Jackson,~K.~A.; Kortus,~J. Interpretation and Automatic Generation of Fermi-Orbital Descriptors. \emph{Journal of Computational Chemistry} \textbf{2019}, \emph{40}, 2843--2857\relax
\mciteBstWouldAddEndPuncttrue
\mciteSetBstMidEndSepPunct{\mcitedefaultmidpunct}
{\mcitedefaultendpunct}{\mcitedefaultseppunct}\relax
\EndOfBibitem
\bibitem[Raghavachari and Trucks(1989)Raghavachari, and Trucks]{raghavachari1989highly}
Raghavachari,~K.; Trucks,~G.~W. Highly correlated systems. Ionization energies of first row transition metals Sc--Zn. \emph{The Journal of Chemical Physics} \textbf{1989}, \emph{91}, 2457--2460\relax
\mciteBstWouldAddEndPuncttrue
\mciteSetBstMidEndSepPunct{\mcitedefaultmidpunct}
{\mcitedefaultendpunct}{\mcitedefaultseppunct}\relax
\EndOfBibitem
\bibitem[Argaman \latin{et~al.}(2013)Argaman, Makov, and Kraisler]{argaman2013higher}
Argaman,~U.; Makov,~G.; Kraisler,~E. Higher ionization energies of atoms in density-functional theory. \emph{Physical Review A} \textbf{2013}, \emph{88}, 042504\relax
\mciteBstWouldAddEndPuncttrue
\mciteSetBstMidEndSepPunct{\mcitedefaultmidpunct}
{\mcitedefaultendpunct}{\mcitedefaultseppunct}\relax
\EndOfBibitem
\bibitem[Kraisler \latin{et~al.}(2010)Kraisler, Makov, and Kelson]{kraisler2010ensemble}
Kraisler,~E.; Makov,~G.; Kelson,~I. Ensemble v-representable ab initio density-functional calculation of energy and spin in atoms: A test of exchange-correlation approximations. \emph{Physical Review A} \textbf{2010}, \emph{82}, 042516\relax
\mciteBstWouldAddEndPuncttrue
\mciteSetBstMidEndSepPunct{\mcitedefaultmidpunct}
{\mcitedefaultendpunct}{\mcitedefaultseppunct}\relax
\EndOfBibitem
\bibitem[Buend{\'\i}a \latin{et~al.}(2013)Buend{\'\i}a, G{\'a}lvez, Maldonado, and Sarsa]{buendia2013quantum}
Buend{\'\i}a,~E.; G{\'a}lvez,~F.; Maldonado,~P.; Sarsa,~A. Quantum Monte Carlo ionization potential and electron affinity for transition metal atoms. \emph{Chemical Physics Letters} \textbf{2013}, \emph{559}, 12--17\relax
\mciteBstWouldAddEndPuncttrue
\mciteSetBstMidEndSepPunct{\mcitedefaultmidpunct}
{\mcitedefaultendpunct}{\mcitedefaultseppunct}\relax
\EndOfBibitem
\bibitem[Johnson and Becke(2017)Johnson, and Becke]{johnson2017communication}
Johnson,~E.~R.; Becke,~A.~D. Communication: DFT treatment of strong correlation in 3d transition-metal diatomics. \emph{The Journal of Chemical Physics} \textbf{2017}, \emph{146}, 211105\relax
\mciteBstWouldAddEndPuncttrue
\mciteSetBstMidEndSepPunct{\mcitedefaultmidpunct}
{\mcitedefaultendpunct}{\mcitedefaultseppunct}\relax
\EndOfBibitem
\bibitem[Furche and Perdew(2006)Furche, and Perdew]{furche2006performance}
Furche,~F.; Perdew,~J.~P. The performance of semilocal and hybrid density functionals in 3d transition-metal chemistry. \emph{The Journal of Chemical Physics} \textbf{2006}, \emph{124}, 044103--044127\relax
\mciteBstWouldAddEndPuncttrue
\mciteSetBstMidEndSepPunct{\mcitedefaultmidpunct}
{\mcitedefaultendpunct}{\mcitedefaultseppunct}\relax
\EndOfBibitem
\bibitem[Holthausen(2005)]{holthausen2005benchmarking}
Holthausen,~M.~C. Benchmarking approximate density functional theory. I. s/d excitation energies in 3d transition metal cations. \emph{Journal of Computational Chemistry} \textbf{2005}, \emph{26}, 1505--1518\relax
\mciteBstWouldAddEndPuncttrue
\mciteSetBstMidEndSepPunct{\mcitedefaultmidpunct}
{\mcitedefaultendpunct}{\mcitedefaultseppunct}\relax
\EndOfBibitem
\bibitem[Harrison(1983)]{harrison1983density}
Harrison,~J.~G. Density functional calculations for atoms in the first transition series. \emph{The Journal of Chemical Physics} \textbf{1983}, \emph{79}, 2265--2269\relax
\mciteBstWouldAddEndPuncttrue
\mciteSetBstMidEndSepPunct{\mcitedefaultmidpunct}
{\mcitedefaultendpunct}{\mcitedefaultseppunct}\relax
\EndOfBibitem
\end{mcitethebibliography}

\pagebreak
\setcounter{table}{0}
\renewcommand{\thetable}{S\arabic{table}}

\section*{Supplemental Information for Atomic Ionization: sd energy imbalance and Perdew-Zunger self-interaction correction energy penalty in 3d atoms}

\noindent
Rohan Maniar, Priyanka B. Shukla, J. Karl Johnson, Koblar A. Jackson, and John P. Perdew

\vspace{1cm}


\begin{table}[h]
\centering
\caption{ Electronic configurations of 3d transition metal atoms (denoted M) and their cations as per the NIST database \cite{NIST_ASD}. Atoms with an asterisk are those that involve the removal of a d-electron in their second ionization.}
\begin{tabular}{ | c | c | c | c | c | c | c | c | c | c | c |}
\hline
 & Atom M & Cation M$^{+}$ & Cation M$^{2+}$ & Cation M$^{3+}$ \\
\hline
Sc & [Ar]3d4s$^{2}$  & [Ar]3d4s  & [Ar]3d  & [Ar] \\
\hline 
Ti & [Ar]3d$^{2}$4s$^{2}$  & [Ar]3d$^{2}$4s  & [Ar]3d$^{2}$ & [Ar]3d  \\
\hline
V$^{*}$ & [Ar]3d$^{3}$4s$^{2}$  & [Ar]3d$^{4}$  & [Ar]3d$^{3}$ & [Ar]3d$^{2}$ \\
\hline
Cr$^{*}$ & [Ar]3d$^{5}$4s  & [Ar]3d$^{5}$  & [Ar]3d$^{4}$ & [Ar]3d$^{3}$ \\
\hline
Mn   & [Ar]3d$^{5}$4s$^{2}$  & [Ar]3d$^{5}$4s  & [Ar]3d$^{5}$ & [Ar]3d$^{4}$  \\
\hline
Fe  & [Ar]3d$^{6}$4s$^{2}$  & [Ar]3d$^{6}$4s  & [Ar]3d$^{6}$ & [Ar]3d$^{5}$ \\ 
\hline
Co$^{*}$ & [Ar]3d$^{7}$4s$^{2}$  & [Ar]3d$^{8}$   & [Ar]3d$^{7}$ & [Ar]3d$^{6}$  \\
\hline
Ni$^{*}$ & [Ar]3d$^{8}$4s$^{2}$  & [Ar]3d$^{9}$  & [Ar]3d$^{8}$ & [Ar]3d$^{7}$  \\
\hline
Cu$^{*}$ & [Ar]3d$^{10}$4s  & [Ar]3d$^{10}$  & [Ar]3d$^{9}$ & [Ar]3d$^{8}$ \\
\hline
Zn & [Ar]3d$^{10}$4s$^{2}$  & [Ar]3d$^{10}$4s  & [Ar]3d$^{10}$ & [Ar]3d$^{9}$ \\
\hline

\end{tabular}

\label{tab: individual electronic configurations}
\end{table}

NOTE: All the following ionization energies have been computed for the DFA or FLOSIC-DFA ground-state electronic configurations, which are not always the experimental configurations shown in Table S1. Appendix A provides an example of how the experimental
ionization was sometimes constructed from the experimental ionization energy from the ground state and an experimental excitation energy.

\begin{table}[t]
\caption{  DFA electronic configurations of 3d transition metal atoms (denoted M) and their cations as per the angular momentum decomposition. For the case of Fe, the reported electronic configuration is that 
for the LSDA or PBE ground state, while r$^{2}$SCAN predicts the NIST electronic configuration ([Ar]3d$^{6}$4s$^{2}$). Atoms with an asterisk are those that involve the removal of a d-electron in their second ionization as per the angular momentum decomposition. Non-NIST electronic configurations are marked in red.}
\centering
\begin{tabular}{ | c | c | c | c | c | c | c | c | c | c | c |}
\hline
 & Atom M & Cation M$^{+}$ & Cation M$^{2+}$ & Cation M$^{+3}$  \\
\hline
Sc & [Ar]3d4s$^{2}$  & [Ar]3d4s  & [Ar]3d & [Ar]  \\
\hline 
Ti$^{*}$ & [Ar]3d$^{2}$4s$^{2}$ & \textcolor{red}{[Ar]3d$^{3}$}  & [Ar]3d$^{2}$ &[Ar]3d  \\
\hline
V$^{*}$ & [Ar]3d$^{3}$4s$^{2}$  & [Ar]3d$^{4}$ & [Ar]3d$^{3}$ & [Ar]3d$^{2}$  \\
\hline
Cr$^{*}$ & [Ar]3d$^{5}$4s  & [Ar]3d$^{5}$  & [Ar]3d$^{4}$ & [Ar]3d$^{3}$ \\
\hline
Mn   & [Ar]3d$^{5}$4s$^{2}$  & [Ar]3d$^{5}$4s  & [Ar]3d$^{5}$ & [Ar]3d$^{4}$ \\
\hline
Fe  & \textcolor{red}{[Ar]3d$^{7}$4s}  & [Ar]3d$^{6}$4s  & [Ar]3d$^{6}$ & [Ar]3d$^{5}$ \\ 
\hline
Co$^{*}$ & \textcolor{red}{[Ar]3d$^{8}$4s} & [Ar]3d$^{8}$   & [Ar]3d$^{7}$ & [Ar]3d$^{6}$ \\
\hline
Ni$^{*}$ & \textcolor{red}{[Ar]3d$^{9}$4s}  & [Ar]3d$^{9}$  & [Ar]3d$^{8}$ & [Ar]3d$^{7}$ \\
\hline
Cu$^{*}$ & [Ar]3d$^{10}$4s  & [Ar]3d$^{10}$  & [Ar]3d$^{9}$ & [Ar]3d$^{8}$\\
\hline
Zn & [Ar]3d$^{10}$4s$^{2}$  & [Ar]3d$^{10}$4s  & [Ar]3d$^{10}$ & [Ar]3d$^{9}$ \\
\hline

\end{tabular}

\label{DFT electronic}
\end{table}

\begin{table}[t]
\caption{ FLOSIC-DFA electronic configurations of 3d transition metal atoms (denoted M) and their cations. Atoms with an asterisk are those that involve the removal of a d-electron in their second ionization as per the angular momentum decomposition (only Cr and Cu). Non-NIST electronic configurations are marked in red. }
\centering
\begin{tabular}{ | c | c | c | c | c | c | c | c | c | c | c |}
\hline
 & Atom M & Cation M$^{+}$ & Cation M$^{2+}$ & Cation M$^{3+}$ \\
\hline
Sc & [Ar]3d4s$^{2}$  & [Ar]3d4s  & [Ar]3d  & [Ar] \\
\hline 
Ti & [Ar]3d$^{2}$4s$^{2}$  & [Ar]3d$^{2}$4s  & [Ar]3d$^{2}$ & [Ar]3d  \\
\hline
V & [Ar]3d$^{3}$4s$^{2}$  & \textcolor{red}{[Ar]3d$^{3}$4s} & [Ar]3d$^{3}$ & [Ar]3d$^{2}$ \\
\hline
Cr$^{*}$ & [Ar]3d$^{5}$4s  & [Ar]3d$^{5}$  & [Ar]3d$^{4}$ & [Ar]3d$^{3}$ \\
\hline
Mn   & [Ar]3d$^{5}$4s$^{2}$  & [Ar]3d$^{5}$4s  & [Ar]3d$^{5}$ & [Ar]3d$^{4}$  \\
\hline
Fe  & [Ar]3d$^{6}$4s$^{2}$  & [Ar]3d$^{6}$4s  & [Ar]3d$^{6}$ & [Ar]3d$^{5}$ \\ 
\hline
Co & [Ar]3d$^{7}$4s$^{2}$  & \textcolor{red}{[Ar]3d$^{7}$4s}   & [Ar]3d$^{7}$ & [Ar]3d$^{6}$  \\
\hline
Ni & [Ar]3d$^{8}$4s$^{2}$  & \textcolor{red}{[Ar]3d$^{8}$4s}  & [Ar]3d$^{8}$ & [Ar]3d$^{7}$  \\
\hline
Cu$^{*}$ & [Ar]3d$^{10}$4s  & [Ar]3d$^{10}$  & [Ar]3d$^{9}$ & [Ar]3d$^{8}$ \\
\hline
Zn & [Ar]3d$^{10}$4s$^{2}$  & [Ar]3d$^{10}$4s  & [Ar]3d$^{10}$ & [Ar]3d$^{9}$ \\
\hline

\end{tabular}
\label{tab:individual electronic configurations}
\end{table}

\begin{table}[]
\begin{tabular}{ccccccc}
\hline
& IP & Expt & Error & $\mid$ Error $\mid$ & \% Error & $\mid$ \% Error $\mid$ \\
\hline
Sc                         & 6.43                 & 6.56                     & -0.13                & 0.13                 & -2.04                & 2.04                 \\
Ti                         & 6.47                 & 6.94                     & -0.47                & 0.47                 & -6.74                & 6.74                 \\
V                          & 6.49                 & 6.75                     & -0.26                & 0.26                 & -3.81                & 3.81                 \\
Cr                         & 7.52                 & 6.77                     & 0.75                 & 0.75                 & 11.09                & 11.09                \\
Mn                         & 7.50                 & 7.43                     & 0.07                 & 0.07                 & 0.87                 & 0.87                 \\
Fe                         & 8.41                 & 7.04                     & 1.37                 & 1.37                 & 19.39                & 19.39                \\
Co                         & 8.09                 & 7.45                     & 0.64                 & 0.64                 & 8.55                 & 8.55                 \\
Ni                         & 8.22                 & 7.61                     & 0.61                 & 0.61                 & 7.95                 & 7.95                 \\
Cu                         & 8.41                 & 7.73                     & 0.67                 & 0.67                 & 8.73                 & 8.73                 \\
Zn                         & 9.69                 & 9.39                     & 0.30                 & 0.30                 & 3.23                 & 3.23                 \\
\hline
Mean  & & & 0.35 & 0.53 & 4.72 & 7.24 \\
\hline
\end{tabular}
\caption{Individual LSDA errors for the first ionization energies of atoms considered. The errors (signed and absolute) are reported in eV.  }
\end{table}

\begin{table}[]
\begin{tabular}{ccccccc}
\hline
& IP & Expt & Error & $\mid$ Error $\mid$ & \% Error & $\mid$ \% Error $\mid$ \\
\hline
Sc                   & 6.29                 & 6.56                 & -0.27                & 0.27                 & -4.09                & 4.09                 \\
Ti                   & 6.28                 & 6.94                 & -0.66                & 0.66                 & -9.51                & 9.51                 \\
V                    & 6.28                 & 6.75                 & -0.47                & 0.47                 & -6.99                & 6.99                 \\
Cr                   & 7.32                 & 6.77                 & 0.55                 & 0.55                 & 8.06                 & 8.06                 \\
Mn                   & 7.16                 & 7.43                 & -0.27                & 0.27                 & -3.62                & 3.62                 \\
Fe                   & 8.11                 & 7.04                 & 1.07                 & 1.07                 & 15.13                & 15.13                \\
Co                   & 7.86                 & 7.45                 & 0.41                 & 0.41                 & 5.50                 & 5.50                 \\
Ni                   & 8.01                 & 7.61                 & 0.40                 & 0.40                 & 5.20                 & 5.20                 \\
Cu                   & 8.16                 & 7.73                 & 0.43                 & 0.43                 & 5.61                 & 5.61                 \\
Zn                   & 9.38                 & 9.39                 & -0.01                & 0.01                 & -0.12                & 0.12                 \\
\hline
Mean  & & & 0.12 &	0.45	& 1.52	& 6.38\\
\hline
\end{tabular}
\caption{Individual PBE errors for the first ionization energies of atoms considered. The errors (signed and absolute) are reported in eV.  }
\end{table}

\begin{table}[]
\begin{tabular}{ccccccc}
\hline
& IP & Expt & Error & $\mid$ Error $\mid$ & \% Error & $\mid$ \% Error $\mid$ \\
\hline
Sc                   & 6.04                 & 6.56                 & -0.52                & 0.52                 & -7.91                & 7.91                 \\
Ti                   & 6.13                 & 6.94                 & -0.81                & 0.81                 & -11.61               & 11.61                \\
V                    & 6.17                 & 6.75                 & -0.58                & 0.58                 & -8.55                & 8.55                 \\
Cr                   & 7.28                 & 6.77                 & 0.51                 & 0.51                 & 7.56                 & 7.56                 \\
Mn                   & 6.88                 & 7.43                 & -0.55                & 0.55                 & -7.43                & 7.43                 \\
Fe                   & 7.75                 & 7.9                  & -0.15                & 0.15                 & -1.94                & 1.94                 \\
Co                   & 7.74                 & 7.45                 & 0.29                 & 0.29                 & 3.83                 & 3.83                 \\
Ni                   & 7.86                 & 7.61                 & 0.25                 & 0.25                 & 3.31                 & 3.31                 \\
Cu                   & 8.05                 & 7.73                 & 0.32                 & 0.32                 & 4.14                 & 4.14                 \\
Zn                   & 9.19                 & 9.39                 & -0.20                & 0.20                 & -2.11                & 2.11                 \\
\hline
Mean  & & & -0.14	& 0.42	& -2.07	& 5.84\\
\hline
\end{tabular}
\caption{Individual r$^{2}$SCAN errors for the first ionization energies of atoms considered. The errors (signed and absolute) are reported in eV.  }
\end{table}

\begin{table}[]
\begin{tabular}{ccccccc}
\hline
& IP & Expt & Error & $\mid$ Error $\mid$ & \% Error & $\mid$ \% Error $\mid$ \\
\hline
Sc                   & 6.49                 & 6.56                 & -0.07                & 0.07                 & -1.02                & 1.02                 \\
Ti                   & 6.71                 & 6.83                 & -0.13                & 0.13                 & -1.83                & 1.83                 \\
V                    & 6.95                 & 7.07                 & -0.12                & 0.12                 & -1.74                & 1.74                 \\
Cr                   & 7.02                 & 6.77                 & 0.25                 & 0.25                 & 3.62                 & 3.62                 \\
Mn                   & 7.35                 & 7.43                 & -0.08                & 0.08                 & -1.08                & 1.08                 \\
Fe                   & 7.82                 & 7.9                  & -0.08                & 0.08                 & -1.04                & 1.04                 \\
Co                   & 8.96                 & 8.3                  & 0.66                 & 0.66                 & 7.96                 & 7.96                 \\
Ni                   & 9.05                 & 8.68                 & 0.37                 & 0.37                 & 4.24                 & 4.24                 \\
Cu                   & 7.79                 & 7.73                 & 0.05                 & 0.05                 & 0.71                 & 0.71                 \\
Zn                   & 9.46                 & 9.39                 & 0.07                 & 0.07                 & 0.70                 & 0.70                 \\
\hline
Mean  & & & 0.09	& 0.19	& 1.05	& 2.39\\
\hline
\end{tabular}
\caption{Individual FLOSIC-LSDA errors for the first ionization energies of atoms considered. The errors (signed and absolute) are reported in eV.  }
\end{table}

\begin{table}[]
\begin{tabular}{ccccccc}
\hline
& IP & Expt & Error & $\mid$ Error $\mid$ & \% Error & $\mid$ \% Error $\mid$ \\
\hline
Sc                   & 6.15                 & 6.56                 & -0.42                & 0.42                 & -6.33                & 6.33                 \\
Ti                   & 6.33                 & 6.83                 & -0.51                & 0.51                 & -7.39                & 7.39                 \\
V                    & 6.54                 & 7.07                 & -0.53                & 0.53                 & -7.51                & 7.51                 \\
Cr                   & 6.14                 & 6.77                 & -0.63                & 0.63                 & -9.25                & 9.25                 \\
Mn                   & 6.86                 & 7.43                 & -0.57                & 0.57                 & -7.74                & 7.74                 \\
Fe                   & 7.33                 & 7.9                  & -0.57                & 0.57                 & -7.20                & 7.20                 \\
Co                   & 8.51                 & 8.3                  & 0.21                 & 0.21                 & 2.52                 & 2.52                 \\
Ni                   & 8.63                 & 8.68                 & -0.05                & 0.05                 & -0.62                & 0.62                 \\
Cu                   & 7.39                 & 7.73                 & -0.35                & 0.35                 & -4.46                & 4.46                 \\
Zn                   & 8.93                 & 9.39                 & -0.46                & 0.46                 & -4.92                & 4.92                 \\
\hline
Mean  & & &-0.39	& 0.43	& -5.29	& 5.79\\
\hline
\end{tabular}
\caption{Individual FLOSIC-PBE errors for the first ionization energies of atoms considered. The errors (signed and absolute) are reported in eV.  }
\end{table}

\begin{table}[]
\begin{tabular}{ccccccc}
\hline
& IP & Expt & Error & $\mid$ Error $\mid$ & \% Error & $\mid$ \% Error $\mid$ \\
\hline
Sc                   & 6.06                 & 6.56                 & -0.50                & 0.50                 & -7.59                & 7.59                 \\
Ti                   & 6.23                 & 6.83                 & -0.60                & 0.60                 & -8.77                & 8.77                 \\
V                    & 6.45                 & 7.07                 & -0.62                & 0.62                 & -8.81                & 8.81                 \\
Cr                   & 6.83                 & 6.77                 & 0.06                 & 0.06                 & 0.84                 & 0.84                 \\
Mn                   & 6.76                 & 7.43                 & -0.68                & 0.68                 & -9.08                & 9.08                 \\
Fe                   & 7.29                 & 7.9                  & -0.61                & 0.61                 & -7.70                & 7.70                 \\
Co                   & 8.47                 & 8.3                  & 0.17                 & 0.17                 & 2.01                 & 2.01                 \\
Ni                   & 8.83                 & 8.68                 & 0.15                 & 0.15                 & 1.71                 & 1.71                 \\
Cu                   & 7.45                 & 7.73                 & -0.28                & 0.28                 & -3.64                & 3.64                 \\
Zn                   & 9.18                 & 9.39                 & -0.21                & 0.21                 & -2.24                & 2.24                 \\
\hline
Mean  & & & -0.31	& 0.39	& -4.33	& 5.24\\
\hline
\end{tabular}
\caption{Individual FLOSIC-r$^{2}$SCAN errors for the first ionization energies of atoms considered. The errors (signed and absolute) are reported in eV.  }
\end{table}

\begin{table}[]
\begin{tabular}{ccccccc}
\hline
& IP & Expt & Error & $\mid$ Error $\mid$ & \% Error & $\mid$ \% Error $\mid$ \\
\hline
Sc                   & 6.16                 & 6.56                 & -0.40                & 0.40                 & -6.08                & 6.08                 \\
Ti                   & 6.36                 & 6.83                 & -0.47                & 0.47                 & -6.90                & 6.90                 \\
V                    & 6.40                 & 7.07                 & -0.67                & 0.67                 & -9.48                & 9.48                 \\
Cr                   & 7.08                 & 6.77                 & 0.31                 & 0.31                 & 4.51                 & 4.51                 \\
Mn                   & 6.80                 & 7.43                 & -0.63                & 0.63                 & -8.44                & 8.44                 \\
Fe                   & 7.62                 & 7.9                  & -0.28                & 0.28                 & -3.56                & 3.56                 \\
Co                   & 8.58                 & 8.3                  & 0.28                 & 0.28                 & 3.37                 & 3.37                 \\
Ni                   & 8.32                 & 8.68                 & -0.36                & 0.36                 & -4.15                & 4.15                 \\
Cu                   & 7.53                 & 7.73                 & -0.20                & 0.20                 & -2.64                & 2.64                 \\
Zn                   & 8.71                 & 9.39                 & -0.68                & 0.68                 & -7.26                & 7.26                 \\
\hline
Mean  & & & -0.31	& 0.43	& -4.06	 & 5.64\\
\hline
\end{tabular}
\caption{Individual one-shot LSIC errors for the first ionization energies of atoms considered. The errors (signed and absolute) are reported in eV.  }
\end{table}

\begin{table}[]
\begin{tabular}{ccccccc}
\hline
& IP & Expt & Error & $\mid$ Error $\mid$ & \% Error & $\mid$ \% Error $\mid$ \\
\hline
Sc                   & 13.43                & 12.8                 & 0.63                 & 0.63                 & 4.88                 & 4.88                 \\
Ti                   & 14.59                & 13.47                & 1.12                 & 1.12                 & 8.34                 & 8.34                 \\
V                    & 15.80                & 14.63                & 1.17                 & 1.17                 & 7.97                 & 7.97                 \\
Cr                   & 17.75                & 16.49                & 1.26                 & 1.26                 & 7.63                 & 7.63                 \\
Mn                   & 16.33                & 15.64                & 0.69                 & 0.69                 & 4.40                 & 4.40                 \\
Fe                   & 16.86                & 16.2                 & 0.66                 & 0.66                 & 4.08                 & 4.08                 \\
Co                   & 18.46                & 17.08                & 1.38                 & 1.38                 & 8.08                 & 8.08                 \\
Ni                   & 19.65                & 18.17                & 1.48                 & 1.48                 & 8.17                 & 8.17                 \\
Cu                   & 21.80                & 20.29                & 1.51                 & 1.51                 & 7.44                 & 7.44                 \\
Zn                   & 18.66                & 17.96                & 0.70                 & 0.70                 & 3.91                 & 3.91                 \\
\hline
Mean  & & & 1.06	& 1.06	& 6.49	& 6.49\\
\hline
\end{tabular}
\caption{Individual LSDA errors for the second ionization energies of atoms considered. The errors (signed and absolute) are reported in eV.  }
\end{table}

\begin{table}[]
\begin{tabular}{ccccccc}
\hline
& IP & Expt & Error & $\mid$ Error $\mid$ & \% Error & $\mid$ \% Error $\mid$ \\
\hline
Sc                   & 13.29                & 12.8                 & 0.49                 & 0.49                 & 3.83                 & 3.83                 \\
Ti                   & 14.38                & 13.47                & 0.91                 & 0.91                 & 6.76                 & 6.76                 \\
V                    & 15.49                & 14.63                & 0.86                 & 0.86                 & 5.85                 & 5.85                 \\
Cr                   & 17.39                & 16.49                & 0.90                 & 0.90                 & 5.48                 & 5.48                 \\
Mn                   & 16.11                & 15.64                & 0.47                 & 0.47                 & 3.02                 & 3.02                 \\
Fe                   & 16.64                & 16.2                 & 0.44                 & 0.44                 & 2.72                 & 2.72                 \\
Co                   & 18.02                & 17.08                & 0.94                 & 0.94                 & 5.48                 & 5.48                 \\
Ni                   & 19.02                & 18.17                & 0.84                 & 0.84                 & 4.65                 & 4.65                 \\
Cu                   & 21.18                & 20.29                & 0.89                 & 0.89                 & 4.40                 & 4.40                 \\
Zn                   & 18.38                & 17.96                & 0.42                 & 0.42                 & 2.32                 & 2.32                 \\
\hline
Mean  & & & 0.72	& 0.72	& 4.45	& 4.45 \\
\hline
\end{tabular}
\caption{Individual PBE errors for the second ionization energies of atoms considered. The errors (signed and absolute) are reported in eV.  }
\end{table}

\begin{table}[]
\begin{tabular}{ccccccc}
\hline
& IP & Expt & Error & $\mid$ Error $\mid$ & \% Error & $\mid$ \% Error $\mid$ \\
\hline
Sc                   & 13.14                & 12.8                 & 0.33                 & 0.33                 & 2.62                 & 2.62                 \\
Ti                   & 14.09                & 13.47                & 0.62                 & 0.62                 & 4.60                 & 4.60                 \\
V                    & 15.06                & 14.63                & 0.43                 & 0.43                 & 2.92                 & 2.92                 \\
Cr                   & 17.06                & 16.49                & 0.57                 & 0.57                 & 3.48                 & 3.48                 \\
Mn                   & 15.97                & 15.64                & 0.33                 & 0.33                 & 2.08                 & 2.08                 \\
Fe                   & 16.48                & 16.2                 & 0.28                 & 0.28                 & 1.73                 & 1.73                 \\
Co                   & 17.51                & 17.08                & 0.43                 & 0.43                 & 2.51                 & 2.51                 \\
Ni                   & 18.14                & 18.17                & -0.03                & 0.03                 & -0.14                & 0.14                 \\
Cu                   & 20.56                & 20.29                & 0.27                 & 0.27                 & 1.32                 & 1.32                 \\
Zn                   & 18.19                & 17.96                & 0.23                 & 0.23                 & 1.27                 & 1.27                 \\
\hline
Mean  & & & 0.35	& 0.35	& 2.24	& 2.27\\
\hline
\end{tabular}
\caption{Individual r$^{2}$SCAN errors for the second ionization energies of atoms considered. The errors (signed and absolute) are reported in eV.  }
\end{table}

\begin{table}[]
\begin{tabular}{ccccccc}
\hline
& IP & Expt & Error & $\mid$ Error $\mid$ & \% Error & $\mid$ \% Error $\mid$ \\
\hline
Sc                   & 12.97                & 12.8                 & 0.16                 & 0.16                 & 1.29                 & 1.29                 \\
Ti                   & 13.63                & 13.58                & 0.05                 & 0.05                 & 0.35                 & 0.35                 \\
V                    & 14.56                & 14.31                & 0.25                 & 0.25                 & 1.76                 & 1.76                 \\
Cr                   & 15.52                & 16.49                & -0.97                & 0.97                 & -5.89                & 5.89                 \\
Mn                   & 15.98                & 15.64                & 0.34                 & 0.34                 & 2.18                 & 2.18                 \\
Fe                   & 16.48                & 16.2                 & 0.28                 & 0.28                 & 1.71                 & 1.71                 \\
Co                   & 16.24                & 16.66                & -0.42                & 0.42                 & -2.51                & 2.51                 \\
Ni                   & 17.00                & 17.13                & -0.13                & 0.13                 & -0.75                & 0.75                 \\
Cu                   & 18.22                & 20.29                & -2.07                & 2.07                 & -10.20               & 10.20                \\
Zn                   & 18.22                & 17.96                & 0.26                 & 0.26                 & 1.44                 & 1.44                 \\
\hline
Mean  & & & -0.22	& 0.49	& -1.06	& 2.81\\
\hline
\end{tabular}
\caption{Individual FLOSIC-LSDA errors for the second ionization energies of atoms considered. The errors (signed and absolute) are reported in eV.  }
\end{table}

\begin{table}[]
\begin{tabular}{ccccccc}
\hline
& IP & Expt & Error & $\mid$ Error $\mid$ & \% Error & $\mid$ \% Error $\mid$ \\
\hline
Sc                   & 12.60                & 12.8                 & -0.20                & 0.20                 & -1.56                & 1.56                 \\
Ti                   & 13.35                & 13.58                & -0.23                & 0.23                 & -1.66                & 1.66                 \\
V                    & 14.12                & 14.31                & -0.20                & 0.20                 & -1.36                & 1.36                 \\
Cr                   & 14.88                & 16.49                & -1.61                & 1.61                 & -9.77                & 9.77                 \\
Mn                   & 15.53                & 15.64                & -0.11                & 0.11                 & -0.72                & 0.72                 \\
Fe                   & 15.99                & 16.2                 & -0.21                & 0.21                 & -1.28                & 1.28                 \\
Co                   & 15.70                & 16.66                & -0.96                & 0.96                 & -5.77                & 5.77                 \\
Ni                   & 16.40                & 17.13                & -0.73                & 0.73                 & -4.29                & 4.29                 \\
Cu                   & 16.56                & 20.29                & -3.74                & 3.74                 & -18.41               & 18.41                \\
Zn                   & 17.67                & 17.96                & -0.29                & 0.29                 & -1.63                & 1.63                 \\
\hline
Mean  & & & -0.83	& 0.83	& -4.65	& 4.65\\
\hline
\end{tabular}
\caption{Individual FLOSIC-PBE errors for the second ionization energies of atoms considered. The errors (signed and absolute) are reported in eV.  }
\end{table}

\begin{table}[]
\begin{tabular}{ccccccc}
\hline
& IP & Expt & Error & $\mid$ Error $\mid$ & \% Error & $\mid$ \% Error $\mid$ \\
\hline
Sc                   & 12.59                & 12.8                 & -0.21                & 0.21                 & -1.64                & 1.64                 \\
Ti                   & 13.41                & 13.58                & -0.17                & 0.17                 & -1.23                & 1.23                 \\
V                    & 14.31                & 14.31                & 0.00                 & 0.00                 & 0.02                 & 0.02                 \\
Cr                   & 15.16                & 16.49                & -1.33                & 1.33                 & -8.05                & 8.05                 \\
Mn                   & 15.62                & 15.64                & -0.02                & 0.02                 & -0.12                & 0.12                 \\
Fe                   & 16.10                & 16.2                 & -0.10                & 0.10                 & -0.64                & 0.64                 \\
Co                   & 15.71                & 16.66                & -0.95                & 0.95                 & -5.68                & 5.68                 \\
Ni                   & 16.19                & 17.13                & -0.95                & 0.95                 & -5.52                & 5.52                 \\
Cu                   & 16.91                & 20.29                & -3.39                & 3.39                 & -16.68               & 16.68                \\
Zn                   & 17.71                & 17.96                & -0.26                & 0.26                 & -1.42                & 1.42                 \\
\hline
Mean  & & & -0.74	& 0.74	& -4.10	& 4.10\\
\hline
\end{tabular}
\caption{Individual FLOSIC-r$^{2}$SCAN errors for the second ionization energies of atoms considered. The errors (signed and absolute) are reported in eV.  }
\end{table}

\begin{table}[]
\begin{tabular}{ccccccc}
\hline
& IP & Expt & Error & $\mid$ Error $\mid$ & \% Error & $\mid$ \% Error $\mid$ \\
\hline
Sc                   & 12.52                & 12.8                 & -0.28                & 0.28                 & -2.16                & 2.16                 \\
Ti                   & 13.27                & 13.58                & -0.31                & 0.31                 & -2.25                & 2.25                 \\
V                    & 14.34                & 14.31                & 0.03                 & 0.03                 & 0.20                 & 0.20                 \\
Cr                   & 16.36                & 16.49                & -0.13                & 0.13                 & -0.81                & 0.81                 \\
Mn                   & 15.78                & 15.64                & 0.14                 & 0.14                 & 0.91                 & 0.91                 \\
Fe                   & 16.06                & 16.2                 & -0.15                & 0.15                 & -0.90                & 0.90                 \\
Co                   & 15.91                & 16.66                & -0.75                & 0.75                 & -4.50                & 4.50                 \\
Ni                   & 16.71                & 17.13                & -0.42                & 0.42                 & -2.43                & 2.43                 \\
Cu                   & 19.92                & 20.29                & -0.37                & 0.37                 & -1.84                & 1.84                 \\
Zn                   & 17.69                & 17.96                & -0.27                & 0.27                 & -1.52                & 1.52                 \\
\hline
Mean  & & & -0.25	& 0.28	& -1.53	& 1.75\\
\hline
\end{tabular}
\caption{Individual one-shot LSIC errors for the second ionization energies of atoms considered. The errors (signed and absolute) are reported in eV.  }
\end{table}

\begin{table}[]
\begin{tabular}{ccccccc}
\hline
& IP & Expt & Error & $\mid$ Error $\mid$ & \% Error & $\mid$ \% Error $\mid$ \\
\hline
Sc & 25.43                & 24.76                & 0.68                 & 0.68                 & 2.74                 & 2.74                 \\
Ti & 28.21                & 27.49                & 0.72                 & 0.72                 & 2.63                 & 2.63                 \\
V  & 30.22                & 29.31                & 0.91                 & 0.91                 & 3.11                 & 3.11                 \\
Cr & 32.14                & 30.96                & 1.18                 & 1.18                 & 3.81                 & 3.81                 \\
Mn & 34.70                & 33.67                & 1.04                 & 1.04                 & 3.08                 & 3.08                 \\
Fe & 31.81                & 30.65                & 1.16                 & 1.16                 & 3.79                 & 3.79                 \\
Co & 34.65                & 33.50                & 1.15                 & 1.15                 & 3.42                 & 3.42                 \\
\hline
\end{tabular}
\caption{Individual LSDA errors for the third ionization energies of atoms considered. The errors (signed and absolute) are reported in eV.  }
\end{table}

\begin{table}[]
\begin{tabular}{ccccccc}
\hline
& IP & Expt & Error & $\mid$ Error $\mid$ & \% Error & $\mid$ \% Error $\mid$ \\
\hline
Sc                   & 25.42                & 24.76                & 0.66                 & 0.66                 & 2.66                 & 2.66                 \\
Ti                   & 28.10                & 27.49                & 0.61                 & 0.61                 & 2.22                 & 2.22                 \\
V                    & 29.99                & 29.31                & 0.68                 & 0.68                 & 2.31                 & 2.31                 \\
Cr                   & 31.74                & 30.96                & 0.78                 & 0.78                 & 2.53                 & 2.53                 \\
Mn                   & 34.31                & 33.67                & 0.64                 & 0.64                 & 1.91                 & 1.91                 \\
Fe                   & 31.64                & 30.65                & 0.99                 & 0.99                 & 3.23                 & 3.23                 \\
Co                   & 34.39                & 33.50                & 0.89                 & 0.89                 & 2.65                 & 2.65                 \\
\hline
\end{tabular}
\caption{Individual PBE errors for the third ionization energies of atoms considered. The errors (signed and absolute) are reported in eV.  }
\end{table}

\begin{table}[]
\begin{tabular}{ccccccc}
\hline
& IP & Expt & Error & $\mid$ Error $\mid$ & \% Error & $\mid$ \% Error $\mid$ \\
\hline
Sc                   & 25.18                & 24.76                & 0.43                 & 0.43                 & 1.72                 & 1.72                 \\
Ti                   & 28.05                & 27.49                & 0.56                 & 0.56                 & 2.02                 & 2.02                 \\
V                    & 29.91                & 29.31                & 0.60                 & 0.60                 & 2.04                 & 2.04                 \\
Cr                   & 31.35                & 30.96                & 0.39                 & 0.39                 & 1.26                 & 1.26                 \\
Mn                   & 34.08                & 33.67                & 0.41                 & 0.41                 & 1.22                 & 1.22                 \\
Fe                   & 30.95                & 30.65                & 0.30                 & 0.30                 & 0.98                 & 0.98                 \\
Co                   & 33.99                & 33.50                & 0.49                 & 0.49                 & 1.47                 & 1.47                 \\
\hline
\end{tabular}
\caption{Individual r$^{2}$SCAN errors for the third ionization energies of atoms considered. The errors (signed and absolute) are reported in eV.  }
\end{table}

\begin{table}[]
\begin{tabular}{ccccccc}
\hline
& IP & Expt & Error & $\mid$ Error $\mid$ & \% Error & $\mid$ \% Error $\mid$ \\
\hline
Sc                   & 26.92                & 24.76                & 2.16                 & 2.16                 & 8.74                 & 8.74                 \\
Ti                   & 29.91                & 27.49                & 2.42                 & 2.42                 & 8.79                 & 8.79                 \\
V                    & 28.88                & 29.31                & -0.43                & 0.43                 & -1.48                & 1.48                 \\
Cr                   & 31.15                & 30.96                & 0.20                 & 0.20                 & 0.63                 & 0.63                 \\
Mn                   & 32.98                & 33.67                & -0.69                & 0.69                 & -2.05                & 2.05                 \\
Fe                   & 32.99                & 30.65                & 2.34                 & 2.34                 & 7.63                 & 7.63                 \\
Co                   & 35.94                & 33.50                & 2.44                 & 2.44                 & 7.30                 & 7.30                 \\
\hline
\end{tabular}
\caption{Individual FLOSIC-LSDA errors for the third ionization energies of atoms considered. The errors (signed and absolute) are reported in eV.  }
\end{table}

\begin{table}[]
\begin{tabular}{ccccccc}
\hline
& IP & Expt & Error & $\mid$ Error $\mid$ & \% Error & $\mid$ \% Error $\mid$ \\
\hline
Sc                   & 25.93                & 24.76                & 1.17                 & 1.17                 & 4.74                 & 4.74                 \\
Ti                   & 28.71                & 27.49                & 1.22                 & 1.22                 & 4.44                 & 4.44                 \\
V                    & 27.51                & 29.31                & -1.80                & 1.80                 & -6.13                & 6.13                 \\
Cr                   & 29.74                & 30.96                & -1.22                & 1.22                 & -3.93                & 3.93                 \\
Mn                   & 31.43                & 33.67                & -2.24                & 2.24                 & -6.66                & 6.66                 \\
Fe                   & 31.76                & 30.65                & 1.11                 & 1.11                 & 3.63                 & 3.63                 \\
Co                   & 34.66                & 33.50                & 1.16                 & 1.16                 & 3.47                 & 3.47                 \\
\hline
\end{tabular}
\caption{Individual FLOSIC-PBE errors for the third ionization energies of atoms considered. The errors (signed and absolute) are reported in eV.  }
\end{table}

\begin{table}[]
\begin{tabular}{ccccccc}
\hline
& IP & Expt & Error & $\mid$ Error $\mid$ & \% Error & $\mid$ \% Error $\mid$ \\
\hline
Sc                   & 25.39                & 24.76                & 0.63                 & 0.63                 & 2.54                 & 2.54                 \\
Ti                   & 28.22                & 27.49                & 0.73                 & 0.73                 & 2.65                 & 2.65                 \\
V                    & 28.16                & 29.31                & -1.16                & 1.16                 & -3.94                & 3.94                 \\
Cr                   & 29.83                & 30.96                & -1.13                & 1.13                 & -3.65                & 3.65                 \\
Mn                   & 32.20                & 33.67                & -1.47                & 1.47                 & -4.38                & 4.38                 \\
Fe                   & 30.73                & 30.65                & 0.08                 & 0.08                 & 0.26                 & 0.26                 \\
Co                   & 33.73                & 33.50                & 0.23                 & 0.23                 & 0.68                 & 0.68                 \\
\hline
\end{tabular}
\caption{Individual FLOSIC-r$^{2}$SCAN errors for the third ionization energies of atoms considered. The errors (signed and absolute) are reported in eV.  }
\end{table}

\begin{table}[]
\begin{tabular}{ccccccc}
\hline
& IP & Expt & Error & $\mid$ Error $\mid$ & \% Error & $\mid$ \% Error $\mid$ \\
\hline
Sc                   & 25.98                & 24.76                & 1.22                 & 1.22                 & 4.92                 & 4.92                 \\
Ti                   & 28.79                & 27.49                & 1.30                 & 1.30                 & 4.73                 & 4.73                 \\
V                    & 29.07                & 29.31                & -0.24                & 0.24                 & -0.82                & 0.82                 \\
Cr                   & 31.54                & 30.96                & 0.58                 & 0.58                 & 1.86                 & 1.86                 \\
Mn                   & 33.60                & 33.67                & -0.07                & 0.07                 & -0.21                & 0.21                 \\
Fe                   & 32.11                & 30.65                & 1.46                 & 1.46                 & 4.76                 & 4.76                 \\
Co                   & 34.83                & 33.50                & 1.33                 & 1.33                 & 3.97                 & 3.97                 \\
\hline
\end{tabular}
\caption{Individual one-shot LSIC-LSDA errors for the third ionization energies of atoms considered. The errors (signed and absolute) are reported in eV.  }
\end{table}

\begin{table}[]
\begin{tabular}{cccccccc}
\hline
 & Ionization& IP & Expt & Error & $\mid$ Error $\mid$ & \% Error & $\mid$ \% Error $\mid$ \\
\hline
Mn & 4th                 & 52.51                & 51.21                & 1.30                 & 1.30                 & 2.54                 & 2.54                 \\
Zn & 4th                 & 61.15                & 59.57                & 1.58                 & 1.58                 & 2.65                 & 2.65                 \\
Ga & 4th                 & 64.62                & 63.24                & 1.38                 & 1.38                 & 2.18                 & 2.18                 \\
Ga & 5th                 & 87.77                & 86.01                & 1.76                 & 1.76                 & 2.05                 & 2.05                 \\
Ge & 4th                 & 46.31                & 45.72                & 0.60                 & 0.60                 & 1.30                 & 1.30                 \\
Ge & 5th                & 91.86                & 90.50                & 1.36                 & 1.36                 & 1.50                 & 1.50                 \\
Ge & 6th                & 117.95               & 115.90               & 2.05                 & 2.05                 & 1.77                 & 1.77                 \\
\hline
\end{tabular}
\caption{Individual LSDA errors for the fourth, fifth, and sixth ionization energies reported in this work. The errors (signed and absolute) are reported in eV. }
\end{table}

\begin{table}[]
\begin{tabular}{cccccccc}
\hline
 & Ionization& IP & Expt & Error & $\mid$ Error $\mid$ & \% Error & $\mid$ \% Error $\mid$ \\
\hline
Mn & 4th                & 52.09                & 51.21                & 0.88                 & 0.88                 & 1.72                 & 1.72                 \\
Zn & 4th                & 60.41                & 59.57                & 0.84                 & 0.84                 & 1.40                 & 1.40                 \\
Ga & 4th                & 63.92                & 63.24                & 0.68                 & 0.68                 & 1.08                 & 1.08                 \\
Ga & 5th                & 87.01                & 86.01                & 1.00                 & 1.00                 & 1.17                 & 1.17                 \\
Ge & 4th               & 46.01                & 45.72                & 0.30                 & 0.30                 & 0.65                 & 0.65                 \\
Ge & 5th               & 91.15                & 90.50                & 0.65                 & 0.65                 & 0.72                 & 0.72                 \\
Ge & 6th                & 117.16               & 115.90               & 1.26                 & 1.26                 & 1.08                 & 1.08                 \\
\hline
\end{tabular}
\caption{Individual PBE errors for the fourth, fifth, and sixth ionization energies reported in this work. The errors (signed and absolute) are reported in eV. }
\end{table}

\begin{table}[]
\begin{tabular}{cccccccc}
\hline
 & Ionization& IP & Expt & Error & $\mid$ Error $\mid$ & \% Error & $\mid$ \% Error $\mid$ \\
\hline
Mn & 4th               & 51.75                & 51.21                & 0.54                 & 0.54                 & 1.06                 & 1.06                 \\
Zn & 4th                 & 59.68                & 59.57                & 0.11                 & 0.11                 & 0.18                 & 0.18                 \\
Ga & 4th                 & 63.43                & 63.24                & 0.19                 & 0.19                 & 0.30                 & 0.30                 \\
Ga & 5th                 & 86.25                & 86.01                & 0.24                 & 0.24                 & 0.28                 & 0.28                 \\
Ge & 4th                 & 45.62                & 45.72                & -0.09                & 0.09                 & -0.20                & 0.20                 \\
Ge & 5th                 & 90.76                & 90.50                & 0.26                 & 0.26                 & 0.28                 & 0.28                 \\
Ge & 6th                 & 116.44               & 115.90               & 0.54                 & 0.54                 & 0.47                 & 0.47                 \\
\hline
\end{tabular}
\caption{Individual r$^{2}$SCAN errors for the fourth, fifth, and sixth ionization energies reported in this work. The errors (signed and absolute) are reported in eV. }
\end{table}

\begin{table}[]
\begin{tabular}{cccccccc}
\hline
 & Ionization& IP & Expt & Error & $\mid$ Error $\mid$ & \% Error & $\mid$ \% Error $\mid$ \\
\hline
Mn & 4th                & 51.97                & 51.21                & 0.76                 & 0.76                 & 1.49                 & 1.49                 \\
Zn & 4th                 & 59.83                & 59.57                & 0.26                 & 0.26                 & 0.44                 & 0.44                 \\
Ga & 4th                 & 62.15                & 63.24                & -1.09                & 1.09                 & -1.72                & 1.72                 \\
Ga & 5th                 & 86.80                & 86.01                & 0.79                 & 0.79                 & 0.92                 & 0.92                 \\
Ge & 4th                 & 46.21                & 45.72                & 0.49                 & 0.49                 & 1.08                 & 1.08                 \\
Ge & 5th                 & 89.81                & 90.50                & -0.69                & 0.69                 & -0.76                & 0.76                 \\
Ge & 6th                 & 117.31               & 115.90               & 1.41                 & 1.41                 & 1.22                 & 1.22                 \\
\hline
\end{tabular}
\caption{Individual FLOSIC-LSDA errors for the fourth, fifth, and sixth ionization energies reported in this work. The errors (signed and absolute) are reported in eV. }
\end{table}

\begin{table}[]
\begin{tabular}{cccccccc}
\hline
 & Ionization& IP & Expt & Error & $\mid$ Error $\mid$ & \% Error & $\mid$ \% Error $\mid$ \\
\hline
Mn & 4th                & 50.23                & 51.21                & -0.98                & 0.98                 & -1.91                & 1.91                 \\
Zn & 4th                & 57.67                & 59.57                & -1.90                & 1.90                 & -3.20                & 3.20                 \\
Ga & 4th                 & 59.85                & 63.24                & -3.39                & 3.39                 & -5.36                & 5.36                 \\
Ga & 5th                 & 84.24                & 86.01                & -1.77                & 1.77                 & -2.06                & 2.06                 \\
Ge & 4th                 & 45.42                & 45.72                & -0.30                & 0.30                 & -0.66                & 0.66                 \\
Ge & 5th                 & 87.13                & 90.50                & -3.37                & 3.37                 & -3.72                & 3.72                 \\
Ge & 6th                 & 114.54               & 115.90               & -1.36                & 1.36                 & -1.17                & 1.17                 \\
\hline
\end{tabular}
\caption{Individual FLOSIC-PBE errors for the fourth, fifth, and sixth ionization energies reported in this work. The errors (signed and absolute) are reported in eV. }
\end{table}

\begin{table}[]
\begin{tabular}{cccccccc}
\hline
 & Ionization& IP & Expt & Error & $\mid$ Error $\mid$ & \% Error & $\mid$ \% Error $\mid$ \\
\hline
Mn & 4th                 & 50.36                & 51.21                & -0.85                & 0.85                 & -1.65                & 1.65                 \\
Zn & 4th                 & 58.17                & 59.57                & -1.41                & 1.41                 & -2.36                & 2.36                 \\
Ga & 4th                 & 60.66                & 63.24                & -2.58                & 2.58                 & -4.09                & 4.09                 \\
Ga & 5th                 & 84.19                & 86.01                & -1.82                & 1.82                 & -2.12                & 2.12                 \\
Ge & 4th                 & 45.28                & 45.72                & -0.43                & 0.43                 & -0.95                & 0.95                 \\
Ge & 5th                 & 88.05                & 90.50                & -2.45                & 2.45                 & -2.71                & 2.71                 \\
Ge & 6th                 & 114.75               & 115.90               & -1.15                & 1.15                 & -0.99                & 0.99                 \\
\hline
\end{tabular}
\caption{Individual FLOSIC-r$^{2}$SCAN errors for the fourth, fifth, and sixth ionization energies reported in this work. The errors (signed and absolute) are reported in eV. }
\end{table}

\begin{table}[]
\begin{tabular}{cccccccc}
\hline
 & Ionization& IP & Expt & Error & $\mid$ Error $\mid$ & \% Error & $\mid$ \% Error $\mid$ \\
\hline
Mn & 4th                 & 52.12                & 51.21                & 0.91                 & 0.91                 & 1.77                 & 1.77                 \\
Zn & 4th                 & 60.13                & 59.57                & 0.56                 & 0.56                 & 0.94                 & 0.94                 \\
Ga & 4th                 & 63.13                & 63.24                & -0.11                & 0.11                 & -0.18                & 0.18                 \\
Ga & 5th                 & 86.80                & 86.01                & 0.79                 & 0.79                 & 0.92                 & 0.92                 \\
Ge & 4th                 & 45.17                & 45.72                & -0.54                & 0.54                 & -1.19                & 1.19                 \\
Ge & 5th                 & 90.53                & 90.50                & 0.03                 & 0.03                 & 0.04                 & 0.04                 \\
Ge & 6th                 & 117.36               & 115.90               & 1.46                 & 1.46                 & 1.26                 & 1.26                 \\
\hline
\end{tabular}
\caption{Individual one-shot LSIC errors for the fourth, fifth, and sixth ionization energies reported in this work. The errors (signed and absolute) are reported in eV. }
\end{table}

\begin{table}[]
\resizebox{\columnwidth}{!}{%
\begin{tabular}{lcccccccc}
\hline
 & LSDA@FLOSIC & \multicolumn{1}{c}{FLOSIC-LSDA} & \multicolumn{1}{l}{LSIC-LSDA} & \multicolumn{1}{c}{FLOSIC-PBE} & \multicolumn{1}{l}{FLOSIC-r$^2$SCAN} & \multicolumn{1}{c}{LSDA} & \multicolumn{1}{c}{PBE} & \multicolumn{1}{c}{r$^2$SCAN} \\
 \hline
Cr$^{+2}$ & 1.019 & -0.967 & -0.129 & -1.607 & -1.324 & 1.263 & 0.907 & 0.579 \\
Mn$^{+3}$ & 0.926 & -0.691 & -0.070 & -2.242 & -1.473 & 1.036 & 0.644 & 0.412 \\
Cu$^{+2}$ & 1.229 & -2.071 & -0.376 & -3.737 & -3.387 & 1.508 & 0.890 & 0.265 \\
Zn$^{+3}$ & 1.236 & -1.464 & 0.006 & -3.445 & -3.343 & 1.388 & 0.766 & 0.215 \\
Ga$^{+4}$ & 1.258 & -1.087 & -0.111 & -3.387 & -2.584 & 1.378 & 0.680 & 0.189 \\
Ge$^{+5}$ & 1.301 & -0.688 & 0.033 & -3.366 & -2.454 & 1.357 & 0.651 & 0.256 \\
\hline
\end{tabular}%
\caption{Individual errors in eV for the removal of the first 3d electron from Cr, Mn, Cu, Zn, Ga and Ge atoms for LSDA, PBE, r$^2$SCAN, LSDA@FLOSIC, FLOSIC-LSDA, FLOSIC-PBE and FLOSIC-r$^2$SCAN functionals. First column represents the final ionized state of the transition metal atoms.}
    \label{tab:3d first}
}
\end{table}

\begin{table}[]
\resizebox{\columnwidth}{!}{%
\begin{tabular}{lcccccccc}
\hline
& LSDA@FLOSIC & FLOSIC-LSDA & \multicolumn{1}{l}{LSIC-LSDA} & FLOSIC-PBE & \multicolumn{1}{l}{FLOSIC-r$^2$SCAN} & LSDA & PBE & r$^2$SCAN \\
\hline
Cr$^{+3}$ & 1.376 & 0.195 & 0.576 & -1.217 & -1.130 & 1.179 & 0.783 & 0.391 \\
Mn$^{+4}$ & 1.605 & 0.761 & 0.908 & -0.980 & -0.846 & 1.303 & 0.879 & 0.542 \\
Cu$^{+3}$ & 1.712 & -0.432 & 0.387 & -2.242 & -1.846 & 1.505 & 0.822 & 0.009 \\
Zn$^{+4}$ & 1.884 & 0.260 & 0.561 & -1.904 & -1.388 & 1.577 & 0.835 & 0.110 \\
Ga$^{+5}$ & 2.193 & 0.792 & 0.794 & -1.769 & -1.820 & 1.762 & 1.003 & 0.243 \\
Ge$^{+6}$ & 2.496 & 1.411 & 1.462 & -1.359 & -1.153 & 2.051 & 1.257 & 0.541 \\
\hline
\end{tabular}%
    \caption{Individual errors in eV for the removal of the second 3d electron from Cr, Mn, Cu, Zn, Ga and Ge atoms for LSDA, PBE, r$^2$SCAN, LSDA@FLOSIC, FLOSIC-LSDA, FLOSIC-PBE and FLOSIC-r$^2$SCAN functionals. First column represents the final ionized state of the transition metal atoms.}
    \label{tab:3d second}
}
\end{table}

\begin{table}[]\centering
    \caption{The second ionization energy (in eV) of the Cu atom computed using the rFLOSIC-LSDA and cFLOSIC-LSDA functionals along with the NIST experimental value.}
\begin{tabular}{ccccc}
\hline
& & rFLOSIC-LSDA & cFLOSIC-LSDA & NIST \\
\hline
Cu$^{+1}$ $\rightarrow$ Cu$^{+2}$ & 3d$^{10}$ $\rightarrow$ 3d$^{9}$  & 18.20 & 19.80 & 20.29 \\
\hline
\end{tabular}
\end{table}
\pagebreak

\clearpage

\end{document}